\documentclass[journal,onecolumn]{IEEEtran}

\usepackage{cite}
\usepackage{url}
\usepackage{graphicx}
\usepackage{xcolor}
\usepackage{listings}
\usepackage{amsmath}
\usepackage{amsthm}
\newtheorem{assumption}{Assumption}
\usepackage[justification=centering]{caption}

\begin{document}
%

\title{CryptoEmu: An Instruction Set Emulator for Computation Over Ciphers}
%

\author{\IEEEauthorblockN{Xiaoyang Gong} \IEEEauthorblockA{xgong35@wisc.edu} \\
\and
\IEEEauthorblockN{Dan Negrut} \IEEEauthorblockA{negrut@wisc.edu}
}

\markboth{CryptoEmu: An Instruction Set Emulator for Computation Over Ciphers}%
{Gong and Negrut: CryptoEmu: An Instruction Set Emulator for Computation Over Ciphers}

\maketitle

\begin{abstract}
 Fully homomorphic encryption (FHE) allows computations over encrypted data. This technique makes privacy-preserving cloud computing a reality. Users can send their encrypted sensitive data to a cloud server, get encrypted results returned and decrypt them, without worrying about data breaches. 
 
 This project report presents a homomorphic instruction set emulator, CryptoEmu, that enables fully homomorphic computation over encrypted data. The software-based instruction set emulator is built upon an open-source, state-of-the-art homomorphic encryption library that supports gate-level homomorphic evaluation. The instruction set architecture supports multiple instructions that belong to the subset of ARMv8 instruction set architecture. The instruction set emulator utilizes parallel computing techniques to emulate every functional unit for minimum latency. This project report includes details on design considerations, instruction set emulator architecture, and datapath and control unit implementation. We evaluated and demonstrated the instruction set emulator's performance and scalability on a 48-core workstation. CryptoEmu has shown a significant speedup in homomorphic computation performance when compared with HELib, a state-of-the-art homomorphic encryption library.
 \end{abstract}


\section{Introduction}
\label{sec:introduction}

\IEEEPARstart{I}{n} the conventional cloud service model, users share data with their
service provide (cloud) to outsource computations. The cloud receives encrypted data and decrypts it with the cloud’s private key or the private key shared between the user and the cloud. Thus, the service provider has access to user data, which might contain sensitive information like health records, bank statements, or trade secrets. Privacy concerns have been raised along with the wide adoption of cloud services. In 2019, over 164.68 million sensitive records were exposed in the United States \cite{Record2019}. 

\begin{figure}[ht]
    \centering {\includegraphics[width=0.6\textwidth]{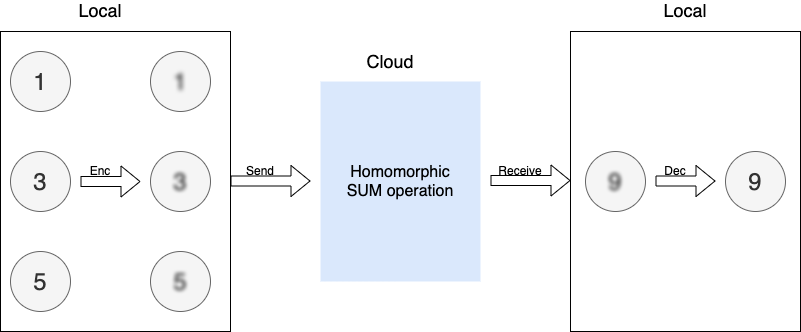}}
    \caption{Homomorphic Encryption}
    \small
    Blurry text in the figure denotes encrypted data.
    \label{f:HEDemo}
\end{figure}

In the worst-case scenario, the cloud service provider cannot be trusted. User data is inherently unsafe if it is in plain text. Even if the service provider is honest, cloud service is prone to fail victims of cybercrime. Security loopholes or sophisticated social engineering attacks expose user privacy on the cloud, and a successful attack usually results in a massive user data leak. One way to eliminate this type of risk is to allow the cloud to operate on the encrypted user data without decrypting it. Fully Homomorphic Encryption (FHE) is a special encryption scheme that allows arbitrary computation over encrypted data without knowing the private key. An FHE enabled cloud service model shown in Fig.~\ref{f:HEDemo}. In this example, the user wants to compute the sum of 1, 3, 5 in the cloud. The user first encrypts data with FHE, then sends the cipher (shown in Fig.~\ref{f:HEDemo} as bubbles with blurry text) to the cloud. When the cloud receives encrypted data, it homomorphically adds all encrypted data together to form an encrypted sum and returns the encrypted sum to the user. The user decrypts the encrypted sum with a secret key, and the result in cleartext is 9 -- the sum of 1, 3, and 5. In the entire process, the cloud has no knowledge of user data input and output. Therefore, user data is safe from the insecure cloud or any attack targeted at the cloud service provider. 

Over the years, the research community has developed various encryption schemes that enable computation over ciphers. TFHE \cite{tfhe2020} is an open-source FHE library that allows fully homomorphic evaluation on arbitrary Boolean circuits. TFHE library supports FHE operations on unlimited numbers of logic gates. Using FHE logic gates provided by TFHE, users can build an application-specific FHE circuit to perform arbitrary computations over encrypted data. While TFHE library has a good performance in gate-by-gate FHE evaluation speed and memory usage \cite{Chillotti2016}, a rigid logic circuit has reusability and scalability issues for general-purpose computing. Also, evaluating a logic circuit in software is slow. Because bitwise FHE operations on ciphers are about one billion times slower than the same operations on plain text, computation time ramps up as the circuit becomes complex.

Herein, we propose a solution that embraces a different approach that draws on a homomorphic instruction set emulator called CryptoEmu. CryptoEmu supports multiple FHE instructions (ADD, SUB, DIV, etc.). When CryptoEmu decodes an instruction, it invokes a pre-built function, referred as functional unit, to perform an FHE operation on input ciphertext. All functional units are built upon FHE gates from TFHE library, and they are accelerated using parallel computing techniques. During execution, the functional units fully utilize a multi-core processor to achieve an optimal speedup. A user would simply reprogram the FHE assembly code for various applications, while relying on the optimized functional units.

This report is organized as follows. Section 2 provides a primer on homomorphic encryption and summarizes related work. Section 3 introduces TFHE, an open-source library for fully homomorphic encryption. TFHE provides the building blocks for CryptoEmu. Section 4 describes CryptoEmu's general architecture. Section 5 and 6 provide detailed instruction set emulator implementations and gives benchmark results on Euler, a CPU/GPU supercomputer. Section 7 analyzes CryptoEmu's scalability and vulnerability, and compared CryptoEmu with a popular FHE software library, HELib \cite{HELib2020}. Conclusions and future directions of investigation/development are provided in Section 8.

\section{Background}
\label{sec:background}

\textbf{Homomorphic Encryption.} Homomorphic encryption (HE) is an encryption scheme that supports computation on encrypted data and generates an encrypted output. When the encrypted output is decrypted, its value is equal to the result when applying equivalent computation on unencrypted data.
HE is formally defined as follows: let $Enc()$ be an HE encryption function, $Dec()$ be an HE decryption function, $f()$ be a function, $g()$ be a homomorphic equivalent of $f()$, and $a$ and $b$ be input data in plaintext. The following equation holds:

\[
f(a, b) = Dec(g(Enc(a), Enc(b))) \; .
\]

An HE scheme is a {\textit{partially homomorphic encryption}} (PHE) scheme if $g()$ supports only either addition or multiplication. An HE scheme is a {\textit{somewhat homomorphic encryption}} (SWHE) scheme if a limited number of $g()$ is allowed to be applied to encrypted data. An HE scheme is a {\textit{fully homomorphic encryption}} (FHE) scheme if any $g()$ can be applied for an unlimited number of times over encrypted data \cite{acar2018survey}. 

The first FHE scheme was proposed by Gentry \cite{Gentry2009}. In HE schemes, the plaintext is encrypted with Gaussian noise. The noise grows after every homomorphic evaluation until the noise becomes too large for the encryption scheme to work. This is the reason that SWHE only allows a limited number of homomorphic evaluations. Gentry introduced a novel technique called ``bootstrapping'' such that a ciphertext can be homomorphically decrypted and homomorphically encrypted with the secret key to reduce Gaussian noise \cite{Gentry2009,chillotti2020tfhe}. Building off \cite{Gentry2009}, \cite{ducas2015fhew} improved bootstrapping to speedup homomorphic evaluations. The TFHE library based on \cite{Chillotti2016} and \cite{chillotti2017faster} is one of the FHE schemes with a fast bootstrapping procedure.

\textbf{Related work.} This project proposed a software-based, multiple-instruction ISA emulator that supports fully homomorphic, general-purpose computation. Several general-purpose HE computer architecture implementations exist in both software and hardware. HELib \cite{halevi2013design} is an FHE software library the implements the Brakerski-Gentry-Vaikuntanathan (BGV) homomorphic encryption scheme \cite{brakerski2014leveled}. HELib supports HE arithmetic such as addition, subtraction, multiplication, and data movement operations. HELib can be treated as an assembly language for general-purpose HE computing. Cryptoleq \cite{mazonka2016cryptoleq} is a software-based one-instruction set computer (OISC) emulator for general-purpose HE computing. Cryptoleq uses Paillier partially homomorphic scheme \cite{paillier1999} and supports Turing-complete SUBLEQ instruction. 
HEROIC \cite{tsoutsos2014heroic} is another OISC architecture implemented on FPGA, based on Paillier partially homomorphic scheme. Cryptoblaze \cite{irena2018cryptoblaze} is a multiple-instruction computer based on non-deterministic Paillier encryption that supports partially homomorphic computation. Cryptoblaze is implemented on the FPGA.

\section{TFHE Library}
\begin{figure}[ht]
    \centering {\includegraphics[width=0.6\textwidth]{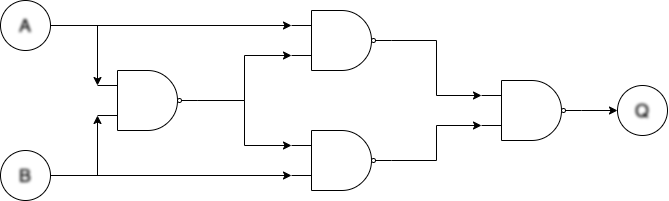}}
    \caption{Use of bootstrapped NAND gate to form arbitrary FHE logic circuit. Blurry text in the figure denotes encrypted data.}
    \label{f:NAND-XOR}
\end{figure}

\
\label{sec:TFHE Library}
TFHE \cite{tfhe2020} is an FHE C/C++ software library used to implement fast gate-by-gate bootstrapping. The idea of TFHE is straightforward: if one can homomorphically evaluate a universal logic gate and homomorphically evaluate the next universal logic gate that uses the previous logic gate's output as its input, one can homomorphically evaluate arbitrary Boolean functions, essentially allowing arbitrary FHE computations on encrypted binary data. Figure~\ref{f:NAND-XOR} demonstrates a minimum FHE gate-level library: NAND gate. Bootstrapped NAND gates are used to construct an FHE XOR gate. Similarly, any FHE logic circuit can be constructed with a combination of bootstrapped NAND gates.

\textbf{TFHE API.} TFHE library contains a comprehensive gate bootstrapping API for the FHE scheme \cite{tfhe2020}, including secret-keyset and cloud-keyset generation; Encryption/decryption with secret-keyset; and FHE evaluation on a binary gate netlist with cloud-keyset. TFHE API's performance is evaluated on a single core of Intel Xeon CPU E5-2650 v3 @ 2.30GHz CPU, running CentOS Linux release 8.2.2004 with 128 GB memory. Table \ref{table:TFHEbenchmark} shows the benchmark result of TFHE APIs that are critical to CryptoEmu's performance. TFHE gate bootstrapping parameter setup, Secret-keyset, and cloud-keyset generation are not included in the table.

\begin{table}[ht]
\begin{center}
\begin{tabular}{llll}
API & Category & Bootstrapped? & Latency (ms) \\ 
\hline
Encrypt & Encrypt decrypt & N/A & 0.0343745  \\ 
Decrypt & Encrypt decrypt & N/A & 0.000319556  \\
\hline
CONSTANT & Homomorphic operations & No & 0.00433995 \\
NOT & Homomorphic operations & No & 0.000679717 \\
COPY & Homomorphic operations & No & 0.000624117 \\
NAND & Homomorphic operations & Yes & 25.5738 \\
OR & Homomorphic operations & Yes & 25.618 \\
AND & Homomorphic operations & Yes & 25.6176 \\
XOR & Homomorphic operations & Yes & 25.6526 \\
XNOR & Homomorphic operations & Yes & 25.795 \\
NOR & Homomorphic operations & Yes & 25.6265 \\
ANDNY & Homomorphic operations & Yes & 25.6982 \\
ANDYN & Homomorphic operations & Yes & 25.684 \\
ORNY & Homomorphic operations & Yes & 25.7787 \\
ORYN & Homomorphic operations & Yes & 25.6957 \\
MUX & Homomorphic operations & Yes & 49.2645 \\
\hline
CreateBitCipher & Ciphertexts & N/A & 0.001725 \\
DeleteBitCipher & Ciphertexts & N/A & 0.002228 \\
ReadBitFromFile & Ciphertexts & N/A & 0.0175304 \\
WriteBitToFile  & Ciphertexts & N/A & 0.00960664
\end{tabular}
\end{center}
\caption{TFHE API Benchmark}
\label{table:TFHEbenchmark}
\end{table}

In Table \ref{table:TFHEbenchmark}, outside the "Homomorphic operations" category, all other operations are relatively fast. In general, the latency is around 25ms, with exceptions of MUX that takes around 50ms, and CONSTANT, NOT, COPY that are relatively fast. The difference in speed is from gate bootstrapping. Unary gates like CONSTANT, NOT and COPY do not need to be bootstrapped. Binary gates need to be bootstrapped once. MUX needs to be bootstrapped twice. The bootstrapping procedure is manifestly the most computationally expensive operation in TFHE. This overhead is alleviated in CryptoEmu via parallel computing as detailed below.

\section{CryptoEmu Architecture Overview}
\begin{figure}[ht]
    \centering {\includegraphics[width=0.7\textwidth]{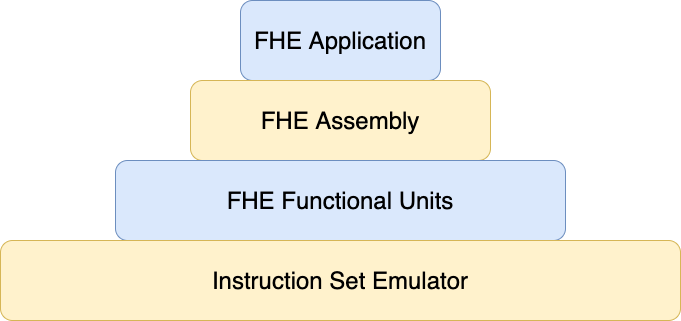}}
    \caption{Abstract Layers}
    \label{f:abstractLayer}
\end{figure}

\
\label{sec: CryptoEmu Architecture Overview}
CryptoEmu is a C/C++ utility that emulates the behavior of Fully Homomorphic Encryption (FHE) instructions. The instruction set that CryptoEmu supports is a subset of ARMv8 A32 instructions for fully homomorphic computation over encrypted data. Figure~\ref{f:abstractLayer} shows the abstract layer for an FHE application. For an FHE application that performs computation over encrypted data, the application will be compiled into FHE assembly that the instruction emulator supports. The instruction set emulator coordinates control units and functional units to decode and execute FHE assembly and returns final results. The design and implementation of CryptoEmu are anchored by two assumptions:

\begin{assumption}
\label{assumption1}
The instruction set emulator runs on a high-performance multi-core machine.
\end{assumption}

\begin{assumption}
\label{assumption2}
The cloud service provider is honest. However, the cloud is subject to cyber-attacks on the user's data.
\end{assumption}

In \S\ref{subsec:Vulnerability} we will discuss modification on CryptoEmu's implementation when Assumption \ref{assumption2} does not hold.

\begin{figure}[ht]
    \centering {\includegraphics[width=0.7\textwidth]{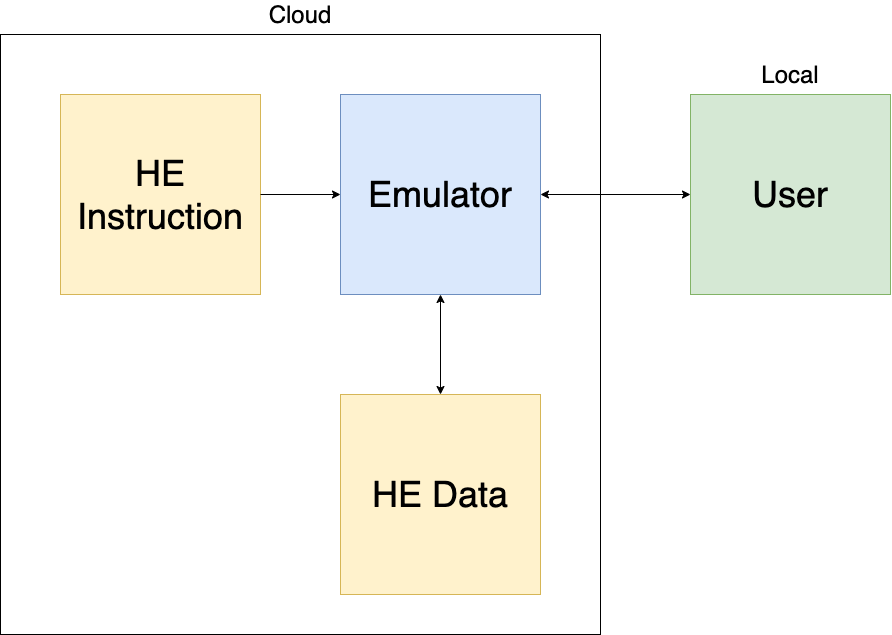}}
    \caption{Cloud service model}
    \label{f:EMUModel}
\end{figure}

\textbf{Cloud service model.} Figure~\ref{f:EMUModel} shows the cloud service model. The instruction set emulator does what an actual hardware asset for encrypted execution would do: it reads from an unencrypted memory space an {\texttt{HE instruction}}; i.e., it fetches instruction that needs to be executed. The instruction set emulator also reads and writes {\texttt{HE data}} from an encrypted memory space, to process the user's data and return encrypted results to the encrypted memory space. The user, or any device that owns the user's secret key, will communicate with the cloud through an encrypted channel. The user provides all encrypted data to cloud. The user can send unencrypted HE instructions to the cloud through a secure channel. The user is also responsible for resolving branch directions for the cloud, based on the encrypted branch taken/non-taken result provided by the cloud. 

\subsection{Data Processing}
\label{subsec:Data processing}
In actuality, the {\texttt{HE instruction}} and {\texttt{HE data}} can be text files or arrays of data bits stored in buffers, if sufficient memory is available. CryptoEmu employs a load-store architecture. All computations occur on virtual registers (vReg), where a vReg is an array of 32 encrypted data bits. Depending on the memory available on the machine, the number of total vReg is configurable. However, it is the compiler's responsibility to recycle vRegs and properly generate read/write addresses. A snippet of possible machine instructions is as follows:

\begin{verbatim}
LOAD    R1  READ_ADDR1
LOAD    R2  READ_ADDR2
ADD     R0  R1, R2
STORE   R0  WRITE_ADDR
\end{verbatim}
Above, to perform a homomorphic addition, CryptoEmu fetches the LOAD instruction from the instruction memory. Because the instruction itself is in cleartext, CryptoEmu decodes the instruction, loads a piece of encrypted data from HE data memory indexed by READ\_ADDR1, and copies the encrypted data into vReg R1. Then, CryptoEmu increments its program counter by 4 bytes, reads the next LOAD instruction, and loads encrypted data from HE data memory into vReg R2. After the two operands are ready, CryptoEmu invokes a 32-bit adder and passes R1, R2 to it. The adder returns encrypted data in R0. Finally, CryptoEmu invokes the STORE operation and writes R0 data into the HE data memory pointed to by WRITE\_ADDR. Under Assumption \ref{assumption2}, the honest cloud could infer some user information from program execution because HE instructions are in cleartext. However, all user data stays encrypted and protected from malicious attackers. Vulnerabilities are discussed in \S \ref{subsec:Vulnerability}.

\subsection{Branch and Control Flow}
\label{subsec:Branch and Control Flow}

CryptoEmu can perform a homomorphic comparison and generate N (negative), Z (zero), C (Unsigned overflow), and V (signed overflow) conditional flags. Based on conditional flags, the branch instruction changes the value of the program counter and therefore changes program flow. Because branches are homomorphically evaluated on the encrypted conditional flag, the branch direction is also encrypted. To solve this problem, CryptoEmu employs a client-server communication model from CryptoBlaze \cite{irena2018cryptoblaze}. Through a secure communication channel, the cloud server will send an encrypted branch decision to a machine (client) that owns the user's private key. The client deciphers the encrypted branch decision and sends the branch decision encrypted with the server's public key to the server. The cloud server finally decrypts the branch decision, and CryptoEmu will move forward with a branch direction. Under assumption \ref{assumption2}, the honest cloud will not use branch decision query and binary search to crack user's encrypted data, nor will the honest cloud infer user information from the user. In \S\ref{subsec:Vulnerability}, the scenario that assumption \ref{assumption2} does not hold will be discussed. 

\section{Data Processing Units}
\label{sec: Data Processing Units}
Data processing units are subroutines that perform manipulation on encrypted data, including homomorphic binary arithmetic, homomorphic bitwise operation, and data movement. Under Assumption \ref{assumption1}, data processing units are implemented with OpenMP \cite{OpenMP} and are designed for parallel computing. If the data processing units exhaust all cores available, the rest of the operations will be serialized. We benchmarked the performance of data processing units with 16-bit and 32-bit vReg size. Benchmarks are based an computing node on Euler. 
The computing node has 2 NUMA nodes. Each NUMA nodes has two sockets, and each socket has a 12-core Intel Xeon CPU E5-2650 v3 @ 2.30GHz CPU. The 48-core computing node runs CentOS Linux release 8.2.2004 with 128 GB memory.

\subsection{Load/Store Unit}
\label{subsec:Load/Store Unit}
CryptoEmu employs a load/store architecture. A LOAD instruction reads data from data memory; a STORE instruction writes data to data memory. The TFHE library \cite{tfhe2020} provides the API for load and store operations on FHE data. If data memory is presented as a file, CryptoEmu invokes the specific LD/ST subroutine, moves the file pointer to the right LD/ST address, and calls the appropriate file IO API, i.e., 
\begin{verbatim}
import_gate_bootstrapping_ciphertext_fromFile()  
\end{verbatim} or 
\begin{verbatim}
export_gate_bootstrapping_ciphertext_toFile()
\end{verbatim}
Preferably, if the machine has available memory, the entire data file is loaded into a buffer as this approach significantly improves LD/ST instruction's performance. Table \ref{table:ldstLatency} shows LD/ST latency for 16-bit and 32-bit. LD/ST on a buffer is significantly faster than LD/ST on a file. The performance speedup is even more when the data file size is large because LD/ST on file needs to use \emph{fseek()} function to access data at the address specified by HE instructions.

\begin{table}[ht]
\begin{center}
\begin{tabular}{|l|l|l|}
\hline
& 16-bit (ms) & 32-bit (ms) \\ 
\hline
Load (file)  & 0.027029 & 0.0554521  \\
\hline
Store (file) & 0.0127804 & 0.0276899  \\
\hline
Load (buffer)  & 0.0043463 & 0.00778488  \\
\hline
Store (buffer) & 0.0043381 & 0.0077692  \\
\hline
\end{tabular}
\end{center}
\caption{LD/ST latencies.}
\label{table:ldstLatency}
\end{table}

\subsection{Adder}
\label{subsec:Adder}
CryptoEmu supports a configurable adder unit of variable width. As for the ISA that CryptoEmu supports, adders are either 16-bit or 32-bit. Operating under an assumption that CryptoEmu runs on a host that supports multi-threading, the adder unit is implemented as a parallel prefix adder \cite{vitoroulis2006parallel}.
The parallel prefix adder has a three-stage structure: pre-calculation of generate and propagate bit; carry propagation; and sum computation. Each stage can be divided into sub-stages and can leverage a multi-core processor. Herein, we use the OpenMP \cite{OpenMP}  library to leverage parallel computing.

\medskip

\noindent \textbf{Stage 1: Propagate and generate calculation.}
Let $a$ and $b$ be the operands to adder, and let $a[i]$ and $b[i]$ be the $i^{th}$ bit of $a$ and $b$. In carry-lookahead logic, $a[i]$ and $b[i]$ generates a carry if $a[i] \ AND \  b[i]$ is 1 and propagates a carry if $a[i] \ XOR \  b[i]$ is 1. This calculation requires an FHE AND gate and an FHE XOR gate, see \S\ref{sec:TFHE Library} and Fig.~\ref{f:NAND-XOR} for gate bootstrapping. An OpenMP parallel region is created to handle two parallel sections. As shown in Fig.~\ref{fig:get_gp}, CryptoEmu spawns two threads to execute two OpenMP sections in parallel.

\begin{figure}
\begin{center}
\begin{verbatim}
#pragma omp parallel sections num_threads(2)
{
    #pragma omp section
    {
        bootsAND(&g_o[0], &a_i[0], &b_i[0], bk);
    }

    #pragma omp section
    {
        bootsXOR(&p_o[0], &a_i[0], &b_i[0], bk);
    }
}
\end{verbatim}
\caption{Parallel optimization for bitwise (g,p) calculation, $ get\_gp() $}
\label{fig:get_gp}
\end{center}
\end{figure}

For a 16-bit adder, \emph{get\_gp()} calculations are applied on every bit. This process is parallelizable: as shown in Fig.~\ref{fig:gp_all}, CryptoEmu spawns 16 parallel sections \cite{OpenMP}, one per bit. Inside each parallel section, the code starts another parallel region that uses two threads. Because of nested parallelism, 32 threads in total are required to calculate every generation and propagate a bit concurrently. If there is an insufficient number of cores, parallel sections will be serialized, which will only affect the efficiency of the emulator. 

\begin{figure}
\begin{center}
\begin{verbatim}
#pragma omp parallel sections num_threads(N)
{
    #pragma omp section
    {
        get_gp(&g[0], &p[0], &a[0], &b[0], bk);
    }
    
    #pragma omp section
    {
        get_gp(&g[1], &p[1], &a[1], &b[1], bk);
    }
    
    ...
    
    #pragma omp section
    {
        get_gp(&g[14], &p[14], &a[14], &b[14], bk);
    }
    
    #pragma omp section
    {
        get_gp(&g[15], &p[15], &a[15], &b[15], bk);
    }
}
\end{verbatim}
\caption{Parallel optimization for (g, p) calculation}
\label{fig:gp_all}
\end{center}

\end{figure}

\medskip

\noindent \textbf{Stage 2: Carry propagation.}
Let $G_{i}$ be the carry signal at $i^{th}$ bit, $P_{i}$ be the accumulated propagation bit at $i^{th}$ bit, $g_{i}$ and $p_{i}$ be outputs from propagate and generate calculation. We define operator $\odot$ such that

\begin{center}
$(g_{x}, p_{x}) \odot (g_{y}, p_{y}) = (g_{x} + p_{x} \cdot g_{y}, p_{x} \cdot p_{y}) \; .$
\end{center}

Carry signal $G_{i}$ and accumulated propagation $P_{i}$ can be recursively defined as

\begin{center}
$(G_{i}, P_{i}) = (g_{i}, p_{i}) \odot (G_{i-1}, P_{i-1})$, where $(G_{0}, P_{0}) = (g_{0}, p_{0}) \; .$
\end{center}

The above recursive formula is equivalent to
\begin{center}
$(G_{i:j}, P_{i:j}) = (G_{i:n}, G_{i:n}) \odot (G_{m:j}, P_{m:j})$, where $i \geq j$ and $m \geq n \; .$
\end{center}

Therefore, carry propagation can be reduced to a parallel scan problem. In CryptoEmu, we defined a routine, \emph{get\_carry()} to perform operation $\odot$. As shown in Fig.~\ref{fig:get_carry} , the $\odot$ requires two FHE AND gate and an FHE OR gate. CryptoEmu spawns two threads to perform the $\odot$ operation in parallel.

\begin{figure}
\begin{center}
\begin{verbatim}
#pragma omp parallel sections num_threads(2)
{
    // Compute carry out (G_i)
    #pragma omp section
    {
        // g(i) = g(i) + p(i) * g(i-1)
        bootsAND(g_tmp, p_1, g_0, bk);
        bootsOR(g_next, g_1, g_tmp, bk);
    }

    #pragma omp section
    {
        // p(i) = p(i) * p(i-1)
        bootsAND(p_next, p_1, p_0, bk);
    }
}
\end{verbatim}
\caption{Parallel optimization for bitwise carry calculation, $ get\_carry() $}
\label{fig:get_carry}
\end{center}
\end{figure}

\begin{figure}[ht]
    \centering {\includegraphics[width=0.7\textwidth]{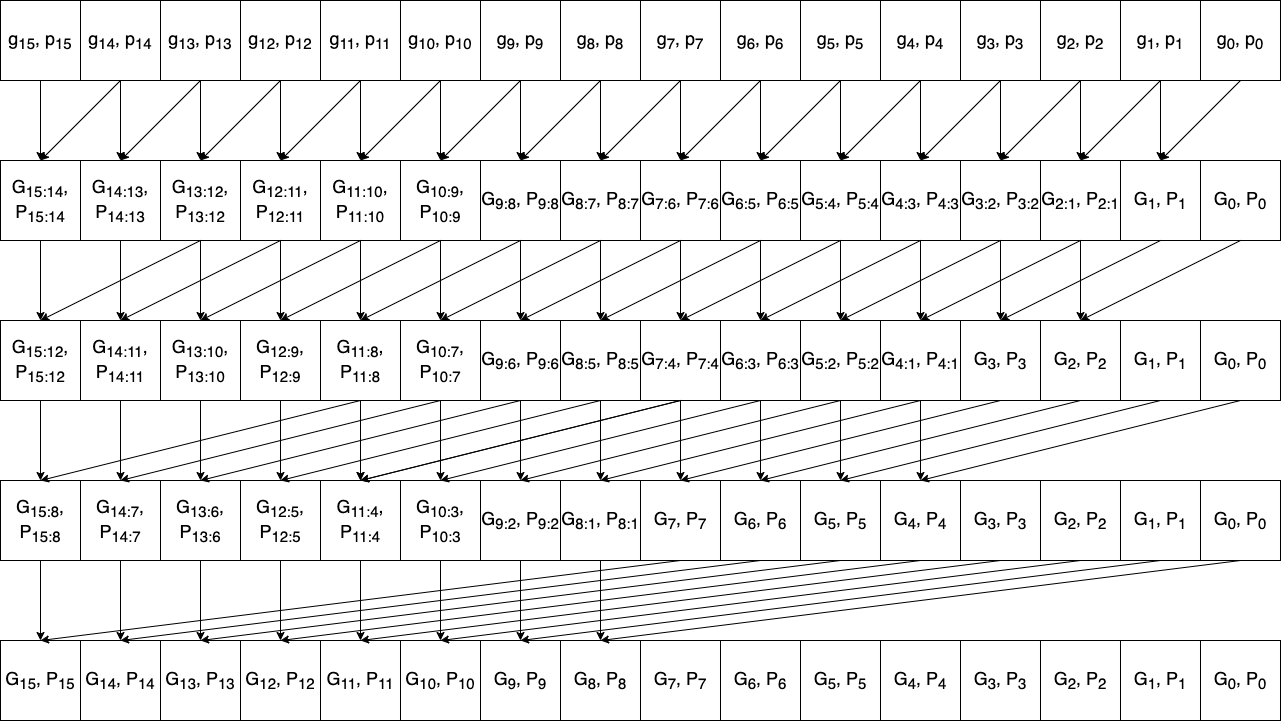}}
    \caption{Parallel scan for carry signals}
    \label{fig:carry_prop}
\end{figure}

For a 16-bit adder, we need 4 levels to compute the carry out from the most significant bit. As shown in fig \ref{fig:carry_prop}, every two arrows that share an arrowhead represents one $\odot$ operation. The $\odot$ operations at the same level can be executed in parallel. In the case of a 16-bit adder, the maximum number of concurrent $\odot$ is 15 at level 1. Because of nested parallelism within the $\odot$ operation, the maximum number of threads required is 30. With a sufficient number of cores, parallel scan reduced carry propagation time from 16 times $\odot$ operation latency, to 4 times $\odot$ operation latency.

\medskip

\noindent \textbf{Stage 3: Sum calculation.} The last stage for parallel prefix adder is sum calculation. Let $s_{i}$ be the sum bit at $i^{th}$ bit, $p_{i}$ be the propagation bit at $i^{th}$ bit,  $G_{i}$ be the carry signal at $i^{th}$ bit. Then

\begin{center}
$s_{i} = p_{i} \ XOR \  G_{i} \; .$
\end{center}

One FHE XOR gate is needed to calculate 1-bit sum. For 16-bit adder, 16 FHE XOR gates are needed. All FHE XOR evaluation are independent, therefore can be executed in parallel. In total 16 threads are required for the best parallel optimization on sum calculation stage.

\paragraph{Benchmark: the 16-bit adder.} Table \ref{table:16BitAdderStages} shows benchmarking results for a 16-bit adder unit executed on the target machine describe earlier in the document. If parallelized, the 1-bit \emph{get\_gp()} shown in Fig.~\ref{fig:get_gp} has one FHE gate latency around 25ms as shown in Table \ref{table:TFHEbenchmark}. Ideally, if sufficient cores are available and there is no overhead from parallel optimization, (g,p) calculation should run 16 \emph{get\_gp()} concurrently, and total latency should be 25ms. In reality, 16-bit (g,p) calculation uses 32 threads and takes 51.39ms to complete due to overhead in parallel computing. 

\begin{table}[ht]
\begin{center}
\begin{tabular}{ll}
Operation & Latency (ms) \\ 
\hline
(g,p) calculation & 51.3939  \\ 
\hline
Carry propagation (Level 1)  & 93.4178  \\
Carry propagation (Level 2)  & 93.5273  \\
Carry propagation (Level 3)  & 80.342   \\
Carry propagation (Level 4)  & 70.8481  \\
\hline
Sum calculation & 34.2846  \\
\hline
Total latency, including overhead & 528.482
\end{tabular}
\end{center}
\caption{16-bit adder latency}
\label{table:16BitAdderStages}
\end{table}

For carry propagation calculation, the 1-bit \emph{get\_carry()} shown in Fig.~\ref{fig:get_carry} has two FHE gate latency of around 50ms when parallelized. In an ideal scenario, each level for carry propagation should run \emph{get\_carry()} in parallel, and total latency should be around 50ms. In reality, the 16-bit carry propagation calculation uses 30 threads on level 1 and takes 93.42ms. A collection of 28 threads are used on carry propagation level 2; the  operation takes 93.58ms. A collection of 24 threads are used on carry propagation level 3; the operation takes 80.34ms. Finally, 16 threads are used on carry propagation level 4; the operation takes 70.85ms.

For sum calculation, 1-bit sum calculation uses one FHE XOR gate, with latency around 25ms. Ideally, if CryptoEmu runs all 16 XOR gates in parallel without parallel computing overhead, the latency for 16-bit sum calculation should be around 25ms. In reality, due to OpenMP overhead, the 16-bit sum calculation uses 16 threads and takes 34.28ms to complete. 

In total, a 16-bit adder's latency is 486.66ms. This result includes latency for all stages, plus overheads like variable declaration, memory allocation, and temporary variable manipulation. 

\paragraph{Benchmark: the 32-bit adder.} Table \ref{table:32BitAdderStages} shows benchmarking results for a 32-bit adder unit executed on the target machine describe earlier in the document. Note that \emph{get\_gp()} and \emph{get\_carry()} have the same performance as the 16-bit adder. If sufficient cores are available, in the absence of OpenMP overhead, the (g,p) calculation should run 32 \emph{get\_gp()} concurrently for 32-bit adder at a total latency of 25ms. In reality, the 32-bit (g,p) calculation uses 32 threads and takes 94.92ms to complete. 
\begin{table}[ht]
\begin{center}
\begin{tabular}{ll}
Operation & Latency (ms) \\ 
\hline
(g,p) calculation & 94.9246  \\ 
\hline
Carry propagation (Level 1)  & 147.451  \\
Carry propagation (Level 2)  & 133.389  \\
Carry propagation (Level 3)  & 127.331  \\
Carry propagation (Level 4)  & 112.268  \\
Carry propagation (Level 5)  & 91.5781  \\
\hline
Sum calculation &  49.0098 \\
\hline
Total latency, including overhead & 941.12
\end{tabular}
\end{center}
\caption{32-bit adder latency}
\label{table:32BitAdderStages}
\end{table}

For carry propagation calculation, if sufficient cores are available, each level for carry propagation should run \emph{get\_carry()} in parallel. Without parallel computing overhead, total latency should be around 50ms. Level 0 of 32-bit carry propagation calculation uses 62 threads. Because the platform on which CryptoEmu is tested has only 48 cores, level 0 carry propagation calculation is serialized and takes 147.45ms to complete. Level 1 carry propagation calculation uses 60 threads, and similar to level 0, its calculation is serialized. Level 1 carry propagation calculation takes 133.39ms. Level 3 carry propagation calculation that uses 56 threads is serialized and takes 127.33ms to complete. Level 4 carry propagation calculation uses 48 threads, and it is possible to run every \emph{get\_carry()} in parallel on our 48-core workstation. Level 4 carry propagation calculation takes 112.27ms. Level 5 carry propagation calculation uses 32 threads. It is able to execute all \emph{get\_carry()} concurrently; level 5 takes 91.58ms to complete.

For sum calculation, the 32-bit adder spawns 32 threads in parallel to perform FHE XOR operation if sufficient cores are available. The latency for the 32-bit sum calculation should be around 25ms. In reality, the 32-bit sum calculation uses 32 threads and takes 49ms to complete. 

In total, 32-bit adder's latency is 941.12ms. This result includes latency for all stages, plus overheads like variable declaration, memory allocation, and temporary variable manipulation.

\subsection{Subtractor}
\label{subsec:Subtractor}
The subtractor unit supports variable ALU size. CryptoEmu supports subtractors with 16-bit operands or 32-bit operands. Let $a$ be the minuend, $b$ be the subtrahend, and $diff$ be the difference. Formula for 2's complement subtraction is:
\begin{center}
$a + NOT(b) + 1 = diff \; .$ 
\end{center}

\begin{figure}[ht]
    \centering {\includegraphics[width=0.5\textwidth]{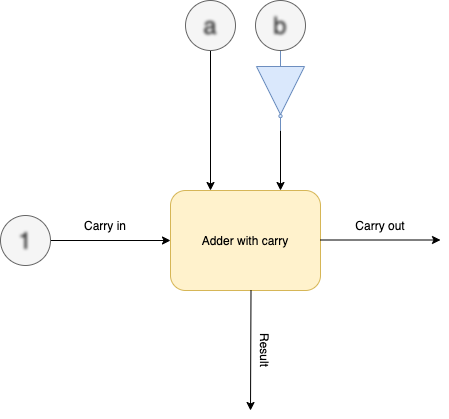}}
    \caption{Subtractor architecture. Blurry text in the figure denotes encrypted data.}
    \label{f:subtractorHE}
\end{figure}

As shown in Fig.~\ref{f:subtractorHE}, CryptoEmu reuses adder units in \S\ref{subsec:Adder} to perform homomorphic subtractions. On the critical path, extra homomorphic NOT gates are used to create subtrahend's complement. For a subtractor with N-bit operands, N homomorphic NOT operations need to be applied on the subtrahend. While all FHE NOT gates can be evaluated in parallel, in \S\ref{sec:TFHE Library} we showed that FHE NOT gates do not need bootstrapping, and is relatively fast (around 0.0007ms latency per gate) comparing to bootstrapped FHE gates. Therefore, parallel execution is not necessary. Instead, the homomorphic NOT operation is implemented in an iterative for-loop, as shown below:
\begin{verbatim}
for(int i = 0; i < N; ++i)
    bootsNOT(&b_neg[i], &b[i], bk);
\end{verbatim}
In addition to homomorphic NOT operation on subtrahend, the carry out bit from bit 0 needs to be evaluated with an OR gate because carry in is 1. Therefore, the adder in \S\ref{subsec:Adder} is extended to take a carry in bit. When sufficient cores are available on the machine, a subtractor units adds negation and carry bit calculation to the adder unit's critical path.

\paragraph{Benchmark.} Tables \ref{table:16BitSubtractor} and \ref{table:32BitSubtractor} report benchmark results for the 16-bit and 32-bit subtractors on the target machine. Negation on the subtrahend takes a trivial amount of time to complete. The homomorphic addition is the most time-consuming operation in the subtractor unit. The homomorphic addition is a little slower than the homomorphic additions in \S\ref{subsec:Adder} because the adder needs to use extra bootstrapped FHE gates to process carry in and calculate carry out from sum bit 0. 

\begin{table}[ht]
\begin{center}
\begin{tabular}{ll}
Operation & Latency (ms) \\ 
\hline
Negation  & 0.0341106               \\ 
Add with carry & 680.59              \\
\hline
Total latency, including overhead & 715.015
\end{tabular}
\end{center}
\caption{16-bit subtractor latency}
\label{table:16BitSubtractor}
\end{table}

\begin{table}[ht]
\begin{center}
\begin{tabular}{ll}
Operation & Latency (ms) \\ 
\hline
Negation  & 0.0347466          \\
Add with carry  & 1058.65           \\
\hline
Total latency, including overhead & 1115.25
\end{tabular}
\end{center}
\caption{32-bit subtractor latency}
\label{table:32BitSubtractor}
\end{table}

\pagebreak 

\subsection{Shifter}
\label{subsec:Shifter}

CryptoEmu supports three types of shifters: logic left shift (LLS), logic right shift (LRS), and arithmetic right shift (ARS). Each shifter type has two modes: immediate and register mode. In immediate mode, the shift amount is in cleartext. For example, the following instruction shifts encrypted data in R0 to left by 1 bit and assigns the shifted value to R0.
\begin{verbatim}
LLS    R0    R0    1 
\end{verbatim}

This instruction is usually issued by the cloud to process user data. Shift immediate implementation is trivial. The shifter calls \emph{bootCOPY()} API to move all data to the input direction by the specified amount. The LSB or MSB will be assigned to an encrypted constant using the \emph{bootCONSTANT()} API call. Because neither \emph{bootCOPY()} nor \emph{bootCONSTANT()} need to be bootstrapped, they are fast operations, see Table \ref{sec:TFHE Library}. Therefore, an iterative loop is used for shifting. Parallel optimization is unnecessary.

In register mode, the shift amount is an encrypted data stored in the input register.
For example, the following instruction shifts encrypted data in R0 to left by the value stored in R1 and assign shifted value to R0.
\begin{verbatim}
LLS    R0    R0    R1
\end{verbatim}

Because the shifting amount stored in R1 is encrypted, the shifter can't simply move all encrypted bits left/right by a certain amount. The shifter is implemented as a barrel shifter, with parallel computing enabled.

\paragraph{Logic  left  shift.} Figure~\ref{f:LLS} shows the architecture for the 16-bit LLS. In the figure, numbers in the bubbles denote encrypted data in the shift register and the shift amount register. Numbers in the diamond denote an encrypted constant value generated by the \emph{bootsCONSTANT()} API. The 16-bit LLS has four stages. In each stage, based on the encrypted shift amount, the FHE MUX homomorphically selects an encrypted bit from the shift register. In the end, the LLS outputs encrypted shifted data. FHE MUX is an elementary logic unit provided by TFHE library \cite{tfhe2020}. FHE MUX needs to be bootstrapped twice, and its latency is around 50ms. Therefore, it is reasonable to spawn multiple threads to execute all MUX select in parallel in each stage. For the 16-bit LLS, each stage needs 16 threads to perform a homomorphic MUX select, as shown in the following code:

\begin{figure}
    \centering {\includegraphics[width=0.485\textwidth]{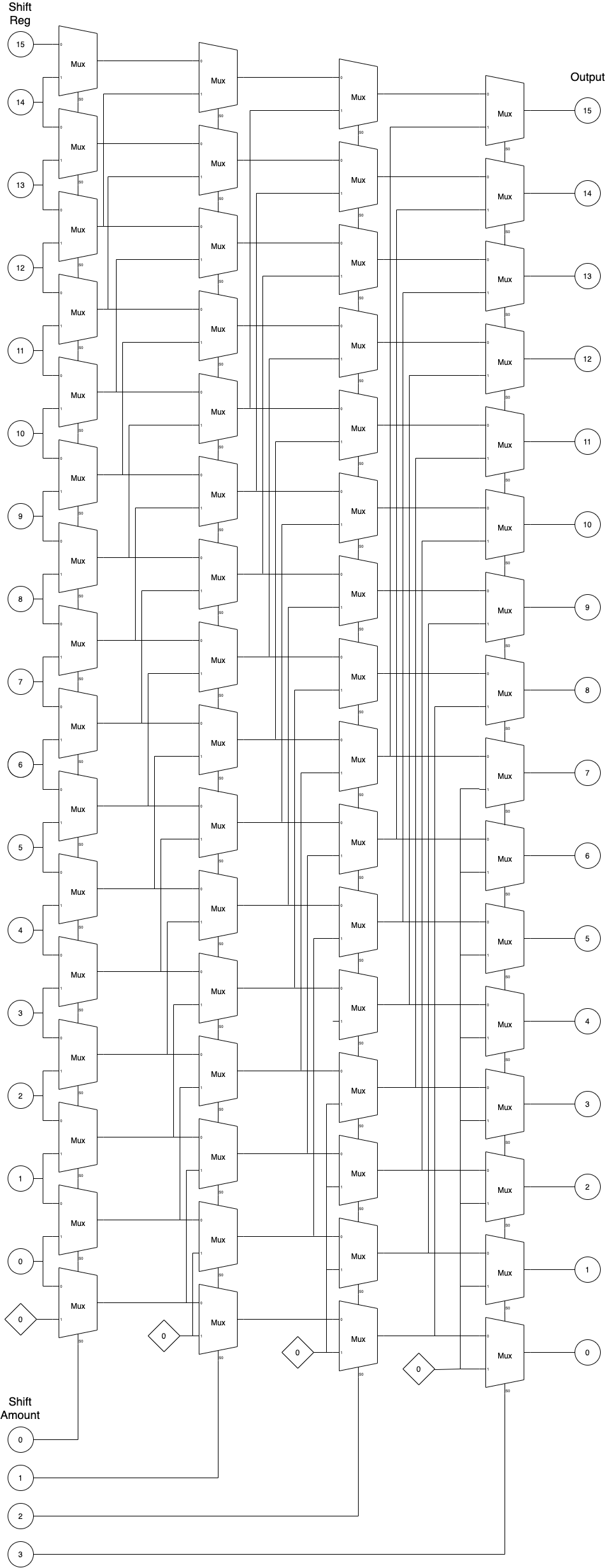}}
    \caption{LLS architecture}
    \label{f:LLS}
\end{figure}

\begin{verbatim}
     #pragma omp parallel sections num_threads(16)
    {
        #pragma omp section
        {
            bootsMUX(&out[0], &amt[0], &zero[0], &in[0], bk);
        }

        #pragma omp section
        {
            bootsMUX(&out[1], &amt[0], &in[0], &in[1], bk);
        }

        ...
        
        #pragma omp section
        {
            bootsMUX(&out[14], &amt[0], &in[13], &in[14], bk);
        }

        #pragma omp section
        {
            bootsMUX(&out[15], &amt[0], &in[14], &in[15], bk);
        }
    }
\end{verbatim}

In a parallel implementation with zero overhead, each stage should have one FHE MUX latency of around 50ms. Therefore, in an ideal scenario, four stages would have a latency of around 200ms.

\paragraph{Benchmark: Logic  left  shift.} Table \ref{table:16BitLLS} shows the benchmark results for the 16-bit LLS on the target platform. Each stage spawns 16 threads to run all FHE MUX in parallel with latency from 50-90ms, a latency that is in between 1 FHE MUX latency to 2 FHE MUX latency, due to parallel computing overhead. In total, it takes around 290ms to carry out a homomorphic LLS operation on 16-bit encrypted data.

\begin{table}[ht]
\begin{center}
\begin{tabular}{ll}
Operation & Latency (ms) \\ 
\hline
Mux select (Stage 1)  & 87.3295  \\
Mux select (Stage 2)  & 82.2656  \\
Mux select (Stage 3)  & 76.6871  \\
Mux select(Stage 4)  & 55.5396  \\
\hline
Total latency, including overhead & 287.92
\end{tabular}
\end{center}
\caption{16-bit LLS latency}
\label{table:16BitLLS}
\end{table}

Table \ref{table:32BitLLS} shows the benchmark result for the 32-bit LLS. Each stage spawns 32 threads and takes around 90-100ms to complete. In total, it takes around 450-500ms for a homomorphic LLS operation on 32-bit encrypted data.

\begin{table} [ht]
\begin{center}
\begin{tabular}{ll}
Operation & Latency (ms) \\ 
\hline
Mux select (Stage 1)  & 101.92  \\
Mux select (Stage 2)  & 89.7695  \\
Mux select (Stage 3)  & 98.8634  \\
Mux select(Stage 4)  & 91.3939  \\
Mux select (Stage 5)  & 91.8491  \\
\hline
Total latency, including overhead & 474.739
\end{tabular}
\end{center}
\caption{32-bit LLS latency}
\label{table:32BitLLS}
\end{table}

\begin{figure}
    \centering {\includegraphics[width=0.485\textwidth]{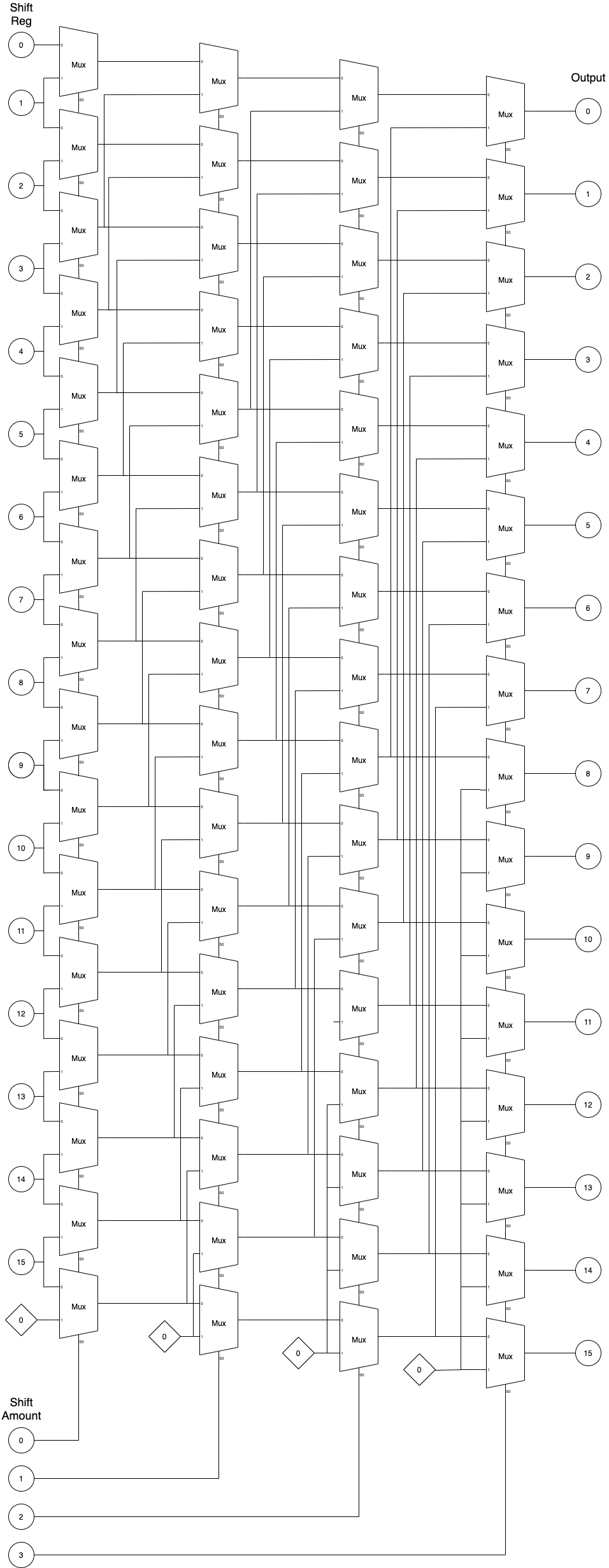}}
    \caption{LRS architecture}
    \label{f:LRS}
\end{figure}

\paragraph{Logic right shift/Arithmetic right shift.} LRS has an architecture that is similar to the architecture of LLS. Figure~\ref{f:LRS} shows the architecture for the 16-bit LRS. Compared to Fig.~\ref{f:LLS}, the only difference between LLS and LRS is the bit order of the input register and output register. To reuse the LRS architecture for ARS, one should simply pass MSB of the shift register as the shift-in value, shown as the numbers in the diamond in Fig.~\ref{f:LRS}. The LRS and ARS shifter implementation is similar to that of LLS. For 16-bit LRS/ARS, CryptoEmu spawns 16 parallel threads to perform a homomorphic MUX select at each stage. 

Table \ref{table:16BitLRS} shows the benchmark result for the 16-bit LRS and ARS. LRS and ARS have similar performance. At each stage, LRS/ARS utilizes 16 threads, and each stage takes 50-90ms to complete. Single stage latency is between 1 FHE MUX latency to 2 FHE MUX latency, due to parallel computing overhead. In total, 16-bit LRS/ARS latency is around 290-300ms.

\begin{table}[ht]
\begin{center}
\begin{tabular}{lll}
Operation & LRS Latency (ms) & ARS Latency (ms)\\ 
\hline
Mux select (Stage 1)  & 88.1408 & 87.7689\\
Mux select (Stage 2)  & 85.5154 & 79.6416\\
Mux select (Stage 3)  & 75.0295 & 75.2938\\
Mux select(Stage 4)  & 55.9639 & 54.9246\\
\hline
Total latency, including overhead & 290.517 & 296.124
\end{tabular}
\end{center}
\caption{16-bit LRS, ARS latency}
\label{table:16BitLRS}
\end{table}

Table \ref{table:32BitLRS} shows the benchmark result for the 32-bit LRS and ARS. The 32-bit LRS/ARS has five stages. Each stage creates 32 parallel threads to evaluate FHE MUX and takes 90-100ms to complete. Single stage latency is around 2 FHE MUX latency. In total, the 32-bit LRS/ARS takes around 470-500ms to complete a homomorphic LRS/ARS operation on 32-bit encrypted data. 

\begin{table}[ht]
\begin{center}
\begin{tabular}{lll}
Operation & LRS Latency (ms) & ARS Latency (ms)\\ 
\hline
Mux select (Level 1)  & 104.33 & 106.132\\
Mux select (Level 2)  & 104.75 & 95.3209\\
Mux select (Level 3)  & 90.2377 & 90.1364\\
Mux select(Level 4)  & 89.7183 & 90.5383\\
Mux select(Level 5)  & 90.6315 & 94.4654\\
\hline
Total latency, including overhead & 472.896 & 491.787
\end{tabular}
\end{center}
\caption{32-bit LRS, ARS latency}
\label{table:32BitLRS}
\end{table}

\subsection{Multiplier}
\label{subsec:Multiplier}

\begin{figure}[ht]
    \centering {\includegraphics[width=0.6\textwidth]{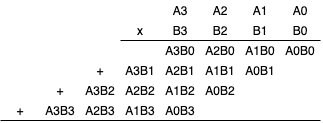}}
    \caption{Binary multiplication}
    \label{f:biMulti}
\end{figure}

\textbf{Design consideration.} Binary multiplication can be treated as a summation of partial products \cite{booth1951signed}, see Fig.~\ref{f:biMulti}. Therefore, adder units mentioned in \S\ref{subsec:Adder} can be reused for partial sum computation. Summing up all partial products is a sum reduction operation, and therefore can be parallelized.

However, the best parallel optimization cannot be achieved on our 48-core computing node. For a 16-bit wide multiplier, the product is a 32-bit encrypted value. Therefore, a 32-bit adder is required to carry out the homomorphic addition. Each 32-bit adder has peak thread usage of 64 threads: 31 threads with nested parallel (g,p) calculation that uses two threads. Thus, on the server used (with 48 cores), the 32-bit adder has to be partially serialized. For the 16-bit multiplier's parallel sum reduction, at most eight summation occur in parallel and each summation uses a 32-bit adder. The peak thread usage is 512 threads. For a 32-bit multiplier, maximum thread usage is 2048 threads. Thus, because homomorphic multiplication is a computationally demanding process, the server used does not have sufficient resources to do all operations in parallel. Homomorphic multiplication will thus show suboptimal performance on the server used in this project.

Based on the design consideration above, CryptoEmu implements a carry-save multiplier \cite{MultiLecture} that supports variable ALU width. Carry-save multiplier uses an adder described in \ref{subsec:Adder} to sum up partial products in series.

\subsubsection{Unsigned multiplication} 
\label{subsubsec:unsignedMultiplication}
\begin{figure}[ht]
    \centering {\includegraphics[width=0.6\textwidth]{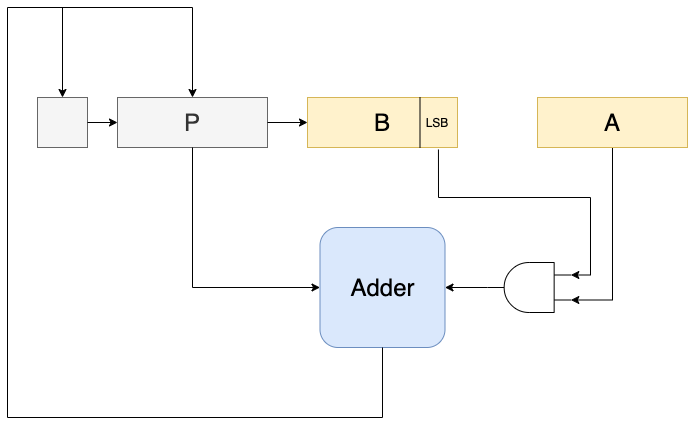}}
    \caption{Carry save multiplier}
    \label{f:carrySave}
\end{figure}

Figure~\ref{f:carrySave} shows the multiplier's architecture. For a 16-bit multiplier, $A$ and $B$ are 16-bit operands stored in vRegs. $P$ is an intermediate 16-bit vReg to store partial products. Adder is a 16-bit adder with a carry in bit, see \S\ref{subsec:Adder}.

On startup, vReg $P$ is initialized to encrypted 0 using TFHE library's bootCONSTANT() API. Next, we enter an iterative loop and homomorphically AND all bits in vReg $A$ with LSB of vReg $B$, and use the result as one of the operands to the 16-bit adder. Data stored in vReg $P$ is then passed to the 16-bit adder as the second operand. The adder performs the homomorphic addition to output an encrypted carry out bit. Next, we right shift carry out, vReg $P$ and vReg $B$, and reached the end of the iterative for-loop. We repeat the for-loop 16 times, and the final product is a 32-bit result stored in vReg $P$ and vReg $B$. The pseudo-code below shows the 16-bit binary multiplication algorithm.
\begin{verbatim}
P = 0;
for 16 times:
    (P, carry out) = P + (B[0] ? A : 0)
    [carry out, P, B] = [carry out, P, B] >> 1;
return [P, B]
\end{verbatim}

For implementation, an N-bit multiplier uses N threads to concurrently evaluate all the FHE AND gates. The adder is already parallel optimized. The rest of the multiplication subroutine is executed sequentially. Therefore, the multiplier is a computationally expensive unit in CryptoEmu. 

\paragraph{Benchmark: Unsigned multiplication.} Table \ref{table:unsignedMulti} shows benchmark for 16-bit unsigned multiplication and 32-bit unsigned multiplication. A single pass for partial product summation takes around 715ms and 925ms, respectively. Total latency is roughly equal to N times single iteration's latency because summation operations are in sequence.
\begin{table}[ht]
\begin{center}
\begin{tabular}{lll}
 & 16-bit multiplier (ms) & 32-bit multiplier (ms)\\ 
\hline
Single iteration  & 715.812 & 926.79\\
Total latency, including overhead & 11316.8 & 36929.2
\end{tabular}
\end{center}
\caption{Unsigned multiplication latency}
\label{table:unsignedMulti}
\end{table}

\subsubsection{Signed multiplication} 
\label{subsubsec:signedMultiplication}
\begin{figure}[ht]
    \centering {\includegraphics[width=0.7\textwidth]{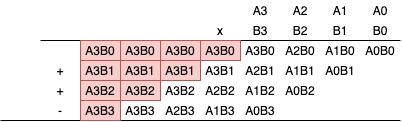}}
    \caption{Signed binary multiplication}
    \label{f:biMultiSigned}
\end{figure}

\begin{figure}[ht]
    \centering {\includegraphics[width=0.7\textwidth]{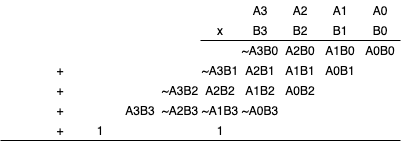}}
    \caption{Simplified signed binary multiplication}
    \label{f:biMultiSignedSim}
\end{figure}

Signed multiplication is implemented using the carry-save multiplier in Figure~\ref{f:carrySave}, with slight modifications. For N-bit signed multiplication, partial products need to be signed extended to 2N bit. Figure~\ref{f:biMultiSigned} shows the partial product summation for 4-bit signed multiplication. This algorithm requires a 2N-bit adder and a 2N-bit subtractor for N-bit signed multiplication. The algorithm is further simplified with a ``magic number'' \cite{MultiLecture}. Figure~\ref{f:biMultiSignedSim} shows the simplified signed multiplication. Based on this algorithm, the unsigned carry-save multiplier is modified to adopt signed multiplication.

\paragraph{Benchmark: Signed multiplication.} The unsigned multiplication was modified for signed multiplication based on the simplified algorithm outlined above. Table \ref{table:signedMulti} shows benchmark results for the 16-bit and 32-bit signed multiplications. A single pass for partial product summation takes roughly 745ms and 1155ms, respectively. The total latency for signed multiplication is around N times that of the single iteration.

\begin{table}[ht]
\begin{center}
\begin{tabular}{lll}
 & 16-bit multiplier (ms) & 32-bit multiplier (ms)\\ 
\hline
Single iteration & 742.8 & 1154.77\\
Total latency, including overhead & 13120.6 & 34826
\end{tabular}
\end{center}
\caption{Signed multiplication latency}
\label{table:signedMulti}
\end{table}

\subsection{Divider}
\label{subsec:Divider}

\begin{figure} [ht]
    \centering {\includegraphics[width=0.5\textwidth]{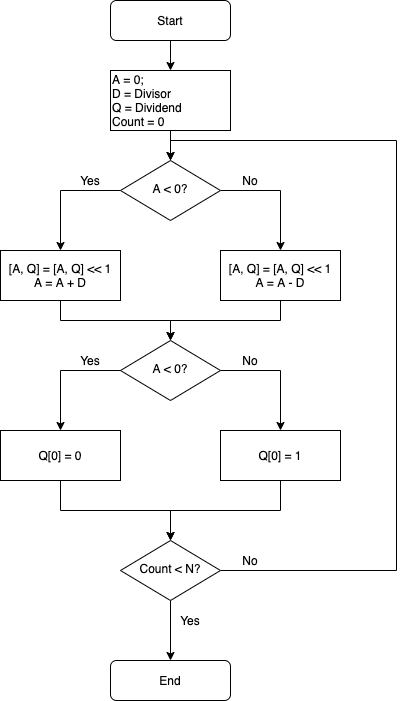}}
    \caption{Binary division algorithm}
    \label{f:divide}
\end{figure}

CryptoEmu supports unsigned division over encrypted data. Figure~\ref{f:divide} shows the flow chart for a non-restoring division algorithm for unsigned integer \cite{Divide2018}.

The division algorithm is based on sequential addition/subtraction. In every iteration, the MSB of vReg $A$ decides whether the divider takes an addition or subtraction. This function is implemented building off the adder in \S\ref{subsec:Adder}. The MSB of vReg $A$ is passed as the $SUB$ bit into the adder. In the adder, the second operand $b$ is homomorphically XORed with the $SUB$ bit and then added with the first operand $A$ and $SUB$ bit. If FHE XOR gates are evaluated in parallel and parallel computing overhead is ignored, the adder with add/sub select is about 1 FHE XOR gate slower than regular adders. Note that, the integer divide instruct does not care about the value of remainder because the result will be rounded down (floor) to the closest integer, and therefore remainder calculation is skipped to save computation time.

The unsigned division algorithm is a sequential algorithm. Within each iteration, a parallel optimized adder subroutine is invoked. Like multiplication, the unsigned division algorithm is computationally expensive. Pseudocode for the 16-bit unsigned division is shown below.

\begin{verbatim}
Non-restoring Division:
    A = 0;
    D = Divisor
    Q = Dividend
    
    for 16 times:
        [A, Q] = [A, Q] << 1
        if A < 0:
            A = A + D
        else:
            A = A - D
        if A < 0:
            Q[0] = 0;
        else:
            Q[0] = 1;
    
    return Q
\end{verbatim}

\paragraph{Benchmark.} Table \ref{table:division} provides benchmark results for the 16-bit and 32-bit unsigned integer divisions. Performance for a single iteration is slightly slower than homomorphic addition because one iteration uses an adder-subtractor unit. The division algorithm is sequential, and the total latency for N-bit division is around N times that of a single iteration's latency.

\begin{table}[ht]
\begin{center}
\begin{tabular}{lll}
 & 16-bit divider (ms) & 32-bit divider (ms)\\ 
\hline
Single iteration & 769.063  &  1308.67\\
Total latency, including overhead & 11659.5 & 38256.5
\end{tabular}
\end{center}
\caption{Unsigned division latency}
\label{table:division}
\end{table}


\subsection{Bitwise operations and data movement}
\label{subsec:BiwiseOperation}
CryptoEmu supports several homomorphic bitwise operations and data movement of configurable data size. The instruction set supports 16-bit and 32-bit bitwise operation and data movement.

The collection of bitwise operations includes homomorphic NOT, AND, OR, XOR, and ORN, in both immediate mode and register mode. All homomorphic bitwise operations except bitwise NOT are implemented with parallel computing optimizations. For the N-bit bitwise operation, CryptoEmu spawns N threads to carry out all FHE gate evaluation in parallel. Bitwise NOT operation is an exception because the FHE NOT gate does not need to be bootstrapped and is relatively fast. Parallel computing cannot be justified because of its overhead. The following code shows a generic way to implement N-bit parallel bitwise operations using the TFHE and OpenMP libraries.

\begin{verbatim}
    #pragma omp parallel sections num_threads(16)
    {
        #pragma omp section
        {
            bootsAND(&out[0], &a[0], &b[0], bk);
        }

        #pragma omp section
        {
            bootsAND(&out[1], &a[1], &b[1], bk);
        }
        
        ...
        
        #pragma omp section
        {
            bootsAND(&out[N-2], &a[N-2], &b[N-2], bk);
        }

        #pragma omp section
        {
            bootsAND(&out[N-1], &a[N-1], &b[N-1], bk);
        }
    }
\end{verbatim}

Data movement instructions support homomorphic operations such as bit field clear, bit field insert, bit-order reverse, and byte-order reverse. Because data movement uses TFHE library's non-bootstrapped APIs like \emph{bootsCOPY()} and \emph{bootsCONSTANT()}, they are not implemented via OpenMP parallel sections.   

\paragraph{Benchmark.} Table \ref{table:bitwiseOps} provides benchmark results for 16-bit and 32-bit bitwise operation and data movement instructions. When parallel optimized, an N-bit bitwise operation spawns N threads. For bitwise operation with bootstrapped FHE gate, latency is between 1-2 FHE gates when all threads are in parallel. The performance of the N-bit bitwise NOT and data movement instructions that are implemented sequentially is proportional to N times the corresponding TFHE API's latency.

\begin{table}[ht]
\begin{center}
\begin{tabular}{lllll}
Operation & Category & Parallel? & 16-bit latency (ms) & 32-bit latency (ms) \\ 
\hline
AND(imm) & Bitwise operations & Yes & 26.9195 & 27.1288\\
AND(reg) & Bitwise operations & Yes & 47.7228 & 55.4597\\
OR(imm) & Bitwise operations & Yes & 28.1923 & 27.8774 \\
OR(reg) & Bitwise operations & Yes & 47.6171 & 50.089 \\
XOR(imm) & Bitwise operations & Yes & 29.2375 & 27.9389\\
XOR(reg) & Bitwise operations & Yes & 47.1986 & 50.3683 \\
ORN(imm) & Bitwise operations & Yes & 27.8467 & 27.776\\
ORN(reg) & Bitwise operations & Yes & 47.6422& 57.0639\\
NOT(imm) & Bitwise operations & Yes & 0.0072794 & 0.0116968\\
NOT(reg) & Bitwise operations & No & 0.006089 & 0.01232\\
\hline
BFC & Bitwise operations & No & 0.0028094 & 0.0026892\\
BFI & Bitwise operations & No & 0.003278 & 0.0033974\\
RBIT & Bitwise operations & No & 0.0104582 & 0.0216222\\
REV & Bitwise operations & No & 0.0097926 & 0.0187618\\
\end{tabular}
\end{center}
\caption{Bitwise operation and data movement latency}
\label{table:bitwiseOps}
\end{table}

\section{Control Units}
\label{sec: Control Units}
A control unit decides and/or changes the value of the program counter (PC), which in CryptoEmu is an integer in cleartext. CryptoEmu supports conditional execution that generates encrypted conditional flags defined in the ARM ISA. Based on conditional flags, the branch unit decides the branch direction to take. The PC is set to a cleartext that points to an HE instruction address when branch is taken, or increased by 4 if the branch is not taken. 

\subsection{Conditional Flags}
\label{subsec:Conditional Flags}
The conditional unit handles NZCV flags defined in the ARM architecture.
The N (negative) flag is set if an instruction's result is negative. 
The Z (zero) flag is set if an instruction's result is zero.
The C (carry) flag is set if an instruction's results in an unsigned overflow.
The V (overflow) flag is set if an instruction's results in an signed overflow.
NZCV values are stored as encrypted data in a vReg. An instruction can be conditionally executed to update NZCV's value. The following code shows HE assembly for a while loop. Note that all data in this example is encrypted.

\begin{verbatim}
C++:
int i = 42;
while(i != 0) 
    i--;
    
HE assembly:
MOV    R0    R0    42
Loop_label:
    SUBS   R0    R0    1
    B_NE   Loop_label
\end{verbatim}

After the SUBS instruction is executed, the NZCV vReg is updated with results in R0. Based on the value of the Z flag, the program counter either updates its value to SUBS' instruction address (branch taken) or increases by 4 (branch non-taken).

Because the homomorphic evaluation of a conditional flag is computationally expensive, an ordinary data processing unit does not have a mechanism to update a conditional flag. In reality, instructions like ADDS, SUBS, and MULS are treated as micro-ops and completed in two steps: homomorphic data processing and homomorphic conditional flag calculation. 

The N flag calculation is straight forward. CryptoEmu takes the MSB of result vReg and assigns it to the NZCV vReg. For the C flag, CryptoEmu takes the carry out result from previous unsigned computation and assigns it to the NZCV vReg. For the Z flag, the conditional unit performs an OR reduction on the input vReg, and assigns a negated result to the NZCV vReg. OR reduction can be parallelized. For a 16-bit conditional unit, OR reduction has four stages and maximum thread usage is eight. For a 32-bit conditional unit, OR reduction has five stages and maximum thread usage is 16. Finally, for the V flag, we take the sign bit (MSB) of two operands and the result for a signed computation, denoted as $a$, $b$, and $s$, respectively. Overflow is then evaluated as
\[
ov = (\Bar{a} \cdot \Bar{b} \cdot s) +  (a \cdot b \cdot \Bar{s}) \; .
\]
Note that the conditional unit calculates overflow in parallel by executing  $(\Bar{a} \cdot \Bar{b} \cdot s)$ and $(a \cdot b \cdot \Bar{s})$ concurrently.

\paragraph{Benchmark.} Table \ref{table:condi} shows benchmark results for the 16-bit and 32-bit conditional units. The N flag and C flag calculations are fast because they do not have bootstrapped FHE gates. The Z flags and V flags are calculated at the same time. For conditional flag calculation on N-bit result, the maximum thread usage is $N/2+2$ threads. Z flag and V flag's latency dominates the total latency. 

\begin{table}[ht]
\begin{center}
\begin{tabular}{lll}
 & 16-bit (ms) & 32-bit  (ms)\\ 
\hline
N flag  & 0.0214161  & 0.0204952 \\
C flag & 0.0007251  & 0.0002539 \\
Z flag and V flag  & 115.811  & 168.728 \\
\hline
Total latency, including overhead & 115.901 & 169.843
\end{tabular}
\end{center}
\caption{Conditional flag latency}
\label{table:condi}
\end{table}

\subsection{Branch}
\label{subsec:Program flow control}

CryptoEmu has a vReg virtual register reserved for the program counter (PC). The PC stores a cleartext value that points at an HE instruction address. CryptoEmu loads an HE instruction from PC, decodes the instruction, invokes the data processing unit to execute the instruction, and repeats. Normally when using the ARM A32 ISA, the next instruction address that CryptoEmu is at the current PC plus 4. However, branch instructions can modify the PC value based on conditional flags, and therefore decide the next instruction address CryptoEmu fetches from. For example, the following code branches to ADDRESS\_LABEL because SUBS instruction sets the Z flag.
\begin{verbatim}
    MOV    R0   R0   1
    SUBS   R0   R0   1    
    B_NE    ADDRESS_LABEL
\end{verbatim}

\begin{figure} [ht]
    \centering {\includegraphics[width=0.6\textwidth]{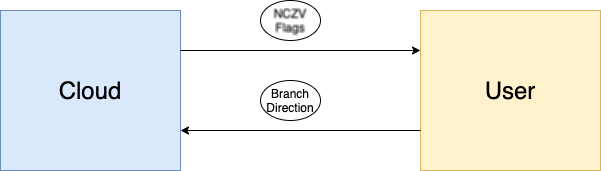}}
    \caption{Server-client model for branch resolution}
    \small
    Blurry text in the figure denotes encrypted data.
    \label{f:branch}
\end{figure}

\S\ref{subsec:Conditional Flags} discussed conditional flag calculation. The NZCV flag is stored in a vReg as encrypted data. The cloud needs to know the value of the NZCV flags to decide which branch direction to take. However, the cloud has no knowledge of NZCV value unless it is decrypted by the user. CryptoEmu adopts a server-client communication model from CryptoBlaze \cite{irena2018cryptoblaze}. As demonstrated in Fig.~\ref{f:branch}, the cloud (server) first homomorphically evaluates an HE instruction and updates the NCZV vReg. Next, the cloud sends the encrypted NCZC value and branch condition to the client. User deciphers the encrypted NCZC value, and sends a branch taken/non-taken decision back to the cloud through a secure channel. Once the cloud learns the branch decision, the PC will either be updated to PC+4, or to the branch address.

\section{Results}\
\label{sec: Results}
This section reports on $(i)$ the scalability with respect to bit count when the core count is a fixed number; $(ii)$ the scalability with respect to core count when the data size is a fixed number; $(iii)$ on CryptoEmu's potential vulnerabilities; and $(iv)$ comparison against the state of the art.

\subsection{Scalability with respect to bit count}
\label{subsec:Scalability with respect to bit count}
Table \ref{table:TFHEbenchmark} shows the latency of the TFHE library API. Non-bootstrapped gates such as CONSTANT, NOT, and COPY are three to four orders of magnitude faster than bootstrapped gates. Therefore, compared to bootstrapped gates, non-bootstrapped gates are considered as negligible in the CryptoEmu's performance equation. Functional units without bootstrapped gates are not included in our scalability analysis.

The latency of a single bootstrapped gate is denoted as $G$. In the rest of the subsection, we analyzed the scalability of functional unit subroutines with respect to data size N. Two scenarios are considered: single core mode, and multi-core mode with a sufficient number of cores to do all intended parallel computation concurrently. CryptoEmu uses all available cores to concurrently perform homomorphic computation on encrypted data with multi-core mode latency. When CryptoEmu exhausted all core resources, the rest of the program will be serialized with single-core mode latency. 

\paragraph{Adder.} An adder has 3 stages: Generate and propagate calculation, carry calculation, and sum calculation. Let the available core number be 1 (single core mode); then, the time complexity for each stage is:

\textit{Generate and propagate calculation}: To compute generate and propagate for one bit, two bootstrapped FHE gates evaluations are required. With N-bit operands, total latency is $2 \cdot G \cdot N$.

\textit{Carry calculation}: Three bootstrapped FHE gates evaluations are required to calculate the carry out bit. With N-bit operands, total latency is $3 \cdot G \cdot N$.

\textit{Sum calculation}: One bootstrapped XOR gate is needed to calculate one sum bit. With N-bit operands, total latency is $G \cdot N$.

\textit{Total latency}: The latency for adder with single core is $6 \cdot G \cdot N$. Time complexity is $O(N)$.

Table \ref{table:adderComplex} summarizes the discussion above for the N-bit adder with one core available.

\begin{table}[ht]
\begin{center}
\begin{tabular}{lll}
 Operation & Latency & Time complexity\\ 
\hline
(g,p) calculation  & $2GN$  & $O(N)$ \\
Carry calculation & $3GN$ & $O(N)$ \\
Sum calculation  & $GN$  & $O(N)$ \\
\hline
Total latency & $6GN$ & $O(N)$
\end{tabular}
\end{center}
\caption{Adder scalability, single core}
\label{table:adderComplex}
\end{table}

Assume a sufficient but fixed number of processors are available. Then, the latencies become:

\textit{Generate and propagate}: Calculation for generate and propagate for one bit can be executed in parallel, with a latency of one bootstrapped FHE gate. Given N-bit operands, N calculations can be carried out concurrently. Therefore, the latency for the (g,p) calculation is $G$.

\textit{Carry calculation}: Carry out computation for one bit can be executed in parallel, with a latency of two bootstrapped FHE gates. For N-bit operands, carry calculation is divided into $log(N)$ stages. Operations in the same stage are executed in parallel. Therefore, latency for carry calculation is $2 \cdot G \cdot log(N)$.

\textit{Sum calculation}: Each sum bit needs one bootstrapped XOR gate. For N-bit operands, N computations can be executed in parallel. Total latency for sum calculation is $G$.

\textit{Total latency}: The total latency for an adder with sufficient yet fixed number of cores is $2 \cdot G \cdot log(N) + 2G$. Time complexity is $O(log(N))$.

Table \ref{table:adderMultiComplex} summarizes the scalability analysis discussion for the N-bit adder with sufficient, yet fixed number of cores available.

\begin{table}[ht]
\begin{center}
\begin{tabular}{lll}
 Operation & Latency & Time complexity\\ 
\hline
(g,p) calculation  & $G$  & $O(1)$ \\
Carry calculation & $2G \cdot log(N)$ & $O(log(N))$ \\
Sum calculation  & $G$  &  $O(1)$ \\
\hline
Total latency & $2G \cdot log(N) + 2G$ & $O(log(N))$
\end{tabular}
\end{center}
\caption{Adder scalability, multiple core}
\label{table:adderMultiComplex}
\end{table}

\paragraph{Subtractor.}
A subtractor performs homomorphic negation on subtrahend, and homomorphically adds subtrahend's complement to minuend using an adder with carry in of one.

Negation uses TFHE library's non-bootstrapped NOT gate, and its latency is negligible. Adding a carry to the adder uses two extra bootstrapped gates. Therefore, the total latency for subtractor is $Latency(Adder) + 2 \cdot G$. 

Table \ref{table:subComplex} summarizes key scalability aspects for the N-bit subtractor with single core and with sufficient yet fixed number of cores available.

\begin{table}[ht]
\begin{center}
\begin{tabular}{lll}
\# of cores & Latency& Time complexity\\ 
\hline
Single core & $6GN + 2G$ & $O(N)$\\
Multi-core & $2Glog(N) + 4G$ & $O(log(N))$\\
\end{tabular}
\end{center}
\caption{Subtractor scalability}
\label{table:subComplex}
\end{table}

\textbf{Shifter.}
For shifting with cleartext immediate, the operation is essentially a homomorphic data movement operation using the TFHE library's COPY and CONSTANT API. Shifting with immediate latency is therefore negligible compared to the rest of function units in CryptoEmu.

Shifting with encrypted vReg is implemented as a barrel shifter. For vReg N-bit shift, the operation has $log(N)$ stages. In each stage, individual bits are selected by bits in the shifting amount vReg moves to the next stage through a MUX. The TFHE MUX has latency around $2 \cdot G$.

When a single core is used, each stage has N MUXs. Latency in one stage is $2 \cdot G \cdot N$. There are $log(N)$ stages, therefore total shifter latency for a single core processor is $2 \cdot G \cdot N \cdot log(N)$.

When sufficient cores are given, MUXs of the same stage can be evaluated in parallel, resulting in a latency of $2 \cdot G$. There are $log(N)$ stages, therefore total shifter latency for processor with sufficient cores is $2 \cdot G \cdot log(N)$.

Table \ref{table:shiftComplex} summarizes key scalability aspects for the shifter with respect to bit count N, in single core and multi-core mode.

\begin{table}[ht]
\begin{center}
\begin{tabular}{lll}
\# of cores & Latency& Time complexity\\ 
\hline
Single core & $2GNlog(N)$ & $O(Nlog(N))$\\
Multi-core & $2Glog(N) $ & $O(log(N))$ \\
\end{tabular}
\end{center}
\caption{Shifter scalability}
\label{table:shiftComplex}
\end{table}

\paragraph{Multiplier.}
The multiplier is implemented with iterative multiplication algorithms. A multiplier with N-bit operands requires N times iterations.

N-bit unsigned multiplication contains an N-bit FHE AND evaluation and invokes an N-bit adder for every iteration. When running unsigned multiplication with a single core, sequential latency for N-bit FHE AND is $N \cdot G$. Sequential latency for the N-bit adder is $6 \cdot G \cdot N$. Latency for one iteration is $7 \cdot G \cdot N$. Total latency for the N-bit unsigned multiplication $7 \cdot G \cdot N^{2}$.

When sufficient number of cores are provided to run subroutines in parallel, parallel latency for the N-bit FHE AND is $G$. Parallel latency for N-bit adder is $2 \cdot G \cdot log(N) + 2G$. One iteration takes $2 \cdot G \cdot log(N) + 3G$ to complete. Total latency for N-bit unsigned multiplication is $2 \cdot G \cdot N \cdot log(N) + 3 \cdot G \cdot N$. 

Signed multiplication has the same latency per iteration as in unsigned multiplication. N-bit signed multiplication involves an N-bit addition that adds an offset to the upper half of the 2N-bit product. This operation has sequential latency of $6 \cdot G \cdot N $ and parallel latency of $2 \cdot G \cdot log(N) + 3G$. Total latency for signed multiplication is therefore $7 \cdot G \cdot N^{2} + 6 \cdot G \cdot N$ for single core, and $2 \cdot G \cdot N \cdot log(N) + 3 \cdot G \cdot N + 2 \cdot G \cdot log(N) + 3G$ if a sufficient number of cores is available.

Table \ref{table:mulComplex} summarizes key scalability aspects (latency and time complexity) for the unsigned and signed multiplication with respect to bit count N, in single core and multi-core mode. 

\begin{table}[ht]
\begin{center}
\begin{tabular}{llll}
unsigned? & \# of cores & Latency& Time complexity\\ 
\hline
Yes & Single core & $7GN^{2}$ & $O(N^{2}))$\\
Yes & Multi-core & $2GNlog(N) + 3GN$  & $O(Nlog(N))$ \\
No & Single core & $7GN^{2} + 6GN$ & $O(N^{2}))$\\
No & Multi-core & $2GNlog(N) + 3GN + 2Glog(N) + 3G$ & $O(Nlog(N))$ \\
\end{tabular}
\end{center}
\caption{Multiplier scalability}
\label{table:mulComplex}
\end{table}

\paragraph{Divider.}
The divider is implemented with iterative non-restoring division algorithm. A divider with N-bit operands requires N iterations. Every iteration involves negligible non-bootstrapped gates and an adder-subtractor unit. The adder-subtractor selectively inverts the second operand based on the value in the SUB bit. Then the subroutine invokes an adder to add the first operand, processes the second operand, and SUB bit together to form a result. 

For computation on single core, selective inversion for an N-bit operand uses N FHE XOR gates, and therefore has a latency of $G \cdot N$. Adder with carry in has $6 \cdot G \cdot N + 2 \cdot G$ latency. Latency per iteration is $7 \cdot G \cdot N + 2 \cdot G$, and total latency is $7 \cdot G \cdot N^{2} + 2 \cdot G \cdot N$ for single core.

If enough cores are available, the N-bit selective inversion can be executed in parallel, resulting in $G$ latency. Adder with carry in parallel computing mode has $2 \cdot G \cdot log(N) + 4 \cdot G$ latency. Latency per iteration is $2 \cdot G \cdot log(N) + 5 \cdot G$, and total latency is $2 \cdot G \cdot N \cdot log(N) + 5 \cdot G \cdot N$.

Table \ref{table:divideComplex} shows latency and time complexity results for the unsigned division with respect to bit count N, in single core and multi-core modes. 

\begin{table}[ht]
\begin{center}
\begin{tabular}{lll}
\# of cores & Latency& Time complexity\\ 
\hline
Single core & $7GN^{2} + 2GN$ & $O(N^{2}))$\\
Multi-core & $2GNlog(N) + 5GN$ & $O(Nlog(N))$ \\
\end{tabular}
\end{center}
\caption{Divider scalability}
\label{table:divideComplex}
\end{table}

\paragraph{Bitwise operations.}
Bitwise operation evaluates all input bits with one FHE gate. Latency per bit is $G$. For N-bit bitwise operation in single core mode, total latency is $G \cdot N$. In multi-core mode, because all FHE gate evaluations are in parallel, the total latency is $G$. 

Table \ref{table:divideComplex} shows latency and time complexity data for bitwise operation with respect to bit count N, in single core and multi-core mode. 

\begin{table}[ht]
\begin{center}
\begin{tabular}{lll}
\# of cores & Latency& Time complexity\\ 
\hline
Single core & $GN$ & $O(N)$\\
Multi-core & $G$ & $O(1)$ \\
\end{tabular}
\end{center}
\caption{Bitwise ops scalability}
\label{table:bitComplex}
\end{table}

\paragraph{Conditional Flags.}
Conditional flags calculation involves four sub-calculations: N flag, Z flag, C flag, and V flag. N flag and C flag calculations are trivial. Z flag calculation contains an OR reduction operation. In single core mode, the N-bit Z flag calculation has a latency of $G \cdot N$. In multi-core mode, the OR reduction is executed in parallel. N-bit OR reduction is broken down into $log(N)$ stages, and in each stage, the FHE OR gates are evaluated in parallel. The N-bit Z flag latency for parallel computing is $Glog(N)$. 

The V flag calculation uses five gates regardless of the size of the input vReg. In single core mode, V flag calculation has $5 \cdot G$ latency. In multi-core mode because V flag calculation can be partially optimized with parallel computing, the latency is $3 \cdot G$. Therefore, total latency for single core mode is $G \cdot N + 5 \cdot G$, and total latency for multi-core mode is  $G \cdot log(N) + 3 \cdot G$.

Table \ref{table:conditionComplex} shows latency and time complexity data for conditional flag calculation with respect to the bit count N, in single core and multi-core mode. 

\begin{table}[ht]
\begin{center}
\begin{tabular}{lll}
\# of cores & Latency& Time complexity\\ 
\hline
Single core & $GN + 5G$ & $O(N)$\\
Multi-core & $Glog(N) + 3G$ & $O(log(N))$ \\
\end{tabular}
\end{center}
\caption{Conditional flags scalability}
\label{table:conditionComplex}
\end{table}

\subsection{Scalability with respect to core count}
\label{subsec:Scalability with respect to core count}

CryptoEmu's performance is dictated by the extent to which it can rely on parallel computing. In ideal scenario, the processor has enough cores to enable any collection of tasks that can be performed in parallel to be processed as such.  Sometimes, a part of the execution of an instruction takes place in parallel, but the rest is sequential because all CPU core resources are exhausted. In the worst case scenario, the processor has only one core and all computations are sequential/serialized.

Assume that the size of the functional units is 16-bit. Peak thread usage is then 32 threads, where each thread performance an FHE gate evaluation. On the server used in this study, which has 48 cores distributed over four CPUs, we vary the number of cores available to carry our emulation from 1 core to 48 cores. This amounts to a strong scaling analysis. We created scenarios where $(i)$ the instruction set can draw on one core only and therefore the computation is sequential; $(ii)$ the instruction set has insufficient cores and parallel computation is mixed with sequential computation; and $(iii)$ the instruction set can count on sufficient cores to do all emulation in parallel.

\begin{figure} [ht]
    \centering {\includegraphics[width=0.6\textwidth]{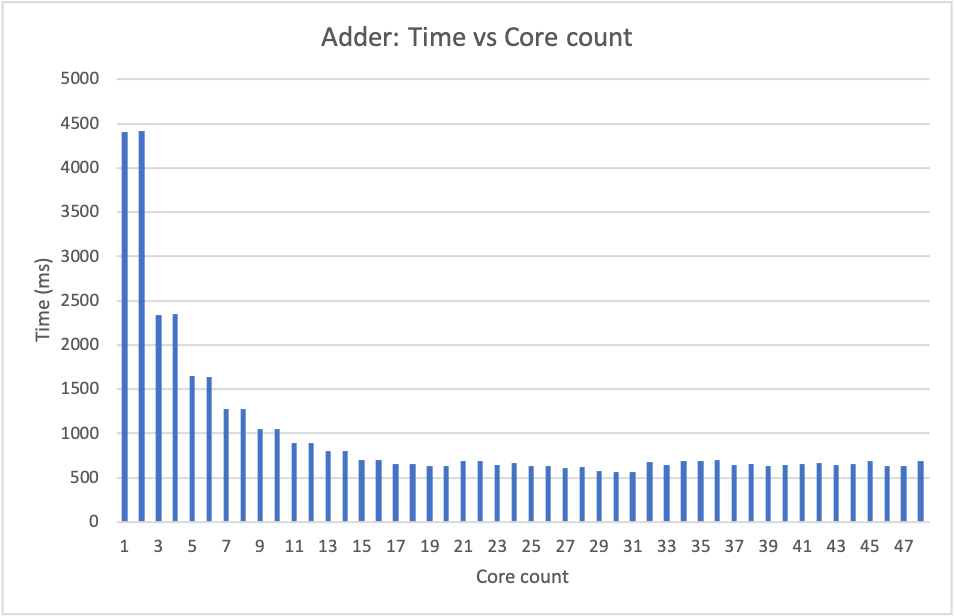}}
    \caption{Time vs Core count}
    \label{f:addercore}
\end{figure}

\paragraph{Adder.} Figure~\ref{f:addercore} shows the 16-bit adder's compute time with respect to core count. From the chart, the adder's latency decreases when the core count increases from 1 to 32. The adder reached its optimal speed at around 32 core count and has a slight decrease in speed when the core count exceeds 32. Adder's maximum speedup from parallel computing is around 7.78x.

\begin{figure} [ht]
    \centering {\includegraphics[width=0.8\textwidth]{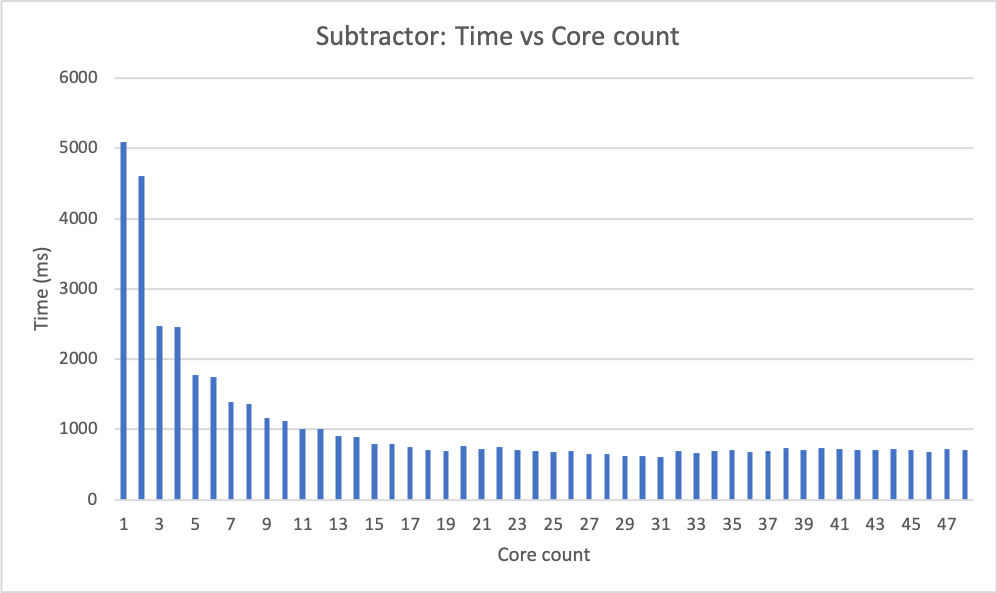}}
    \caption{Time vs Core count}
    \label{f:subcore}
\end{figure}

\paragraph{Subtractor.} Figure~\ref{f:subcore} shows the 16-bit subtractor's compute time with respect to core count. The 16-bit subtractor uses at most 32 cores to compute in parallel. The subtractor's latency decreases as the core count increases from 1-32, and reaches its maximum performance at around 32 core count. At and beyond 32 core count, all parallel optimizations are enabled. When the core counts saturate the subtractor's parallel computing requirement, the latency of the subtractor is around 700ms. The subtractor's maximum speedup from parallel computing is about 8.49x.

\begin{figure} [ht]
    \centering {\includegraphics[width=0.8\textwidth]{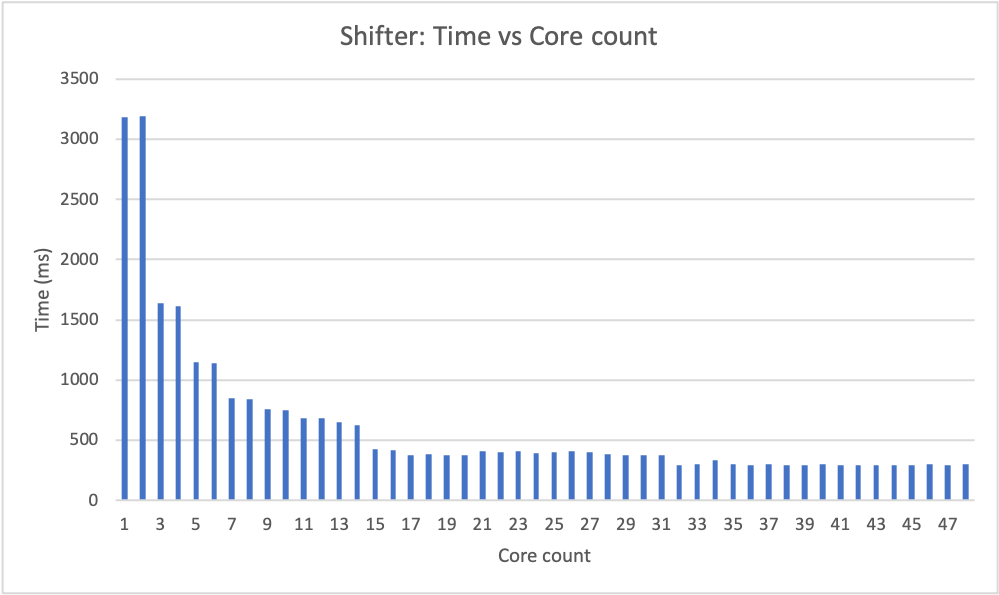}}
    \caption{Time vs Core count}
    \label{f:shiftcore}
\end{figure}

\paragraph{Shifter.} Figure~\ref{f:shiftcore} shows the 16-bit shifter's compute time with respect to core count. A 16-bit shifter has a maximum core usage of 16. The 16-bit shifter's latency decreases as the core count increases from 1-16. The shifter's performance reaches its peak at core count 16. For core counts higher than 16, its stays stable at around 300ms. The maximum speedup yielded by parallel computing is around 10.94x.

\begin{figure} [ht]
    \centering {\includegraphics[width=0.85\textwidth]{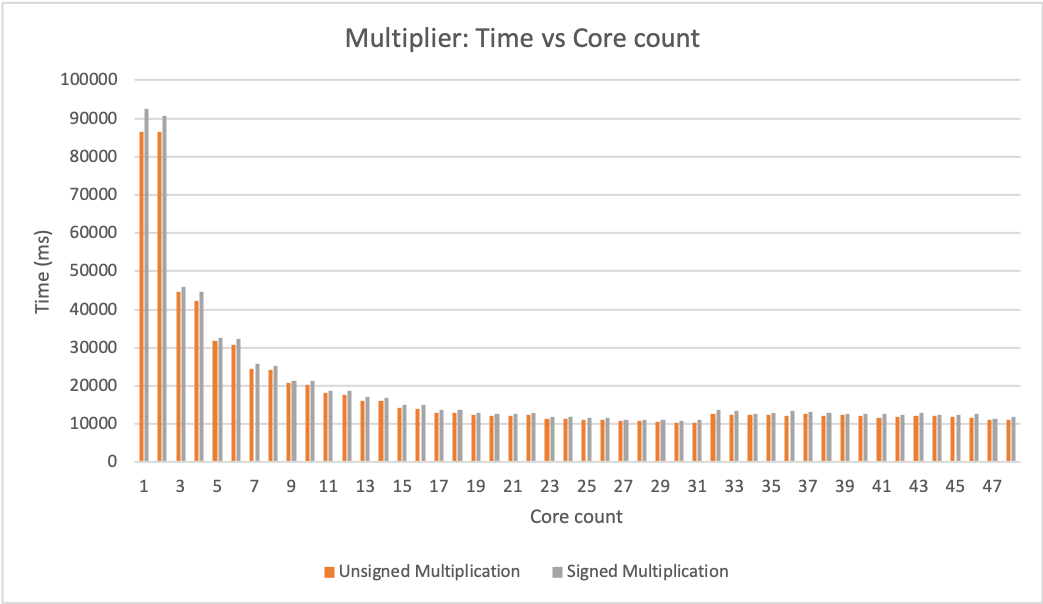}}
    \caption{Time vs Core count}
    \label{f:multicore}
\end{figure}

\paragraph{Multiplier.} Figure~\ref{f:multicore} shows the 16-bit multiplier's compute time with respect to core count. A 16-bit multiplier requires 32 cores to enable all parallel computing optimization. From the chart, both unsigned and signed multiplication's latency decrease as the number of cores increases from 1 to 32. The multiplier reached its top speed at around 32 cores. Beyond that, the multiplier's latency stays around 11000-12000ms. Unsigned multiplication has a maximum parallel computing speedup of 8.3x; signed multiplication has a maximum parallel computing speedup of 8.4x.

\begin{figure} [ht]
    \centering {\includegraphics[width=0.8\textwidth]{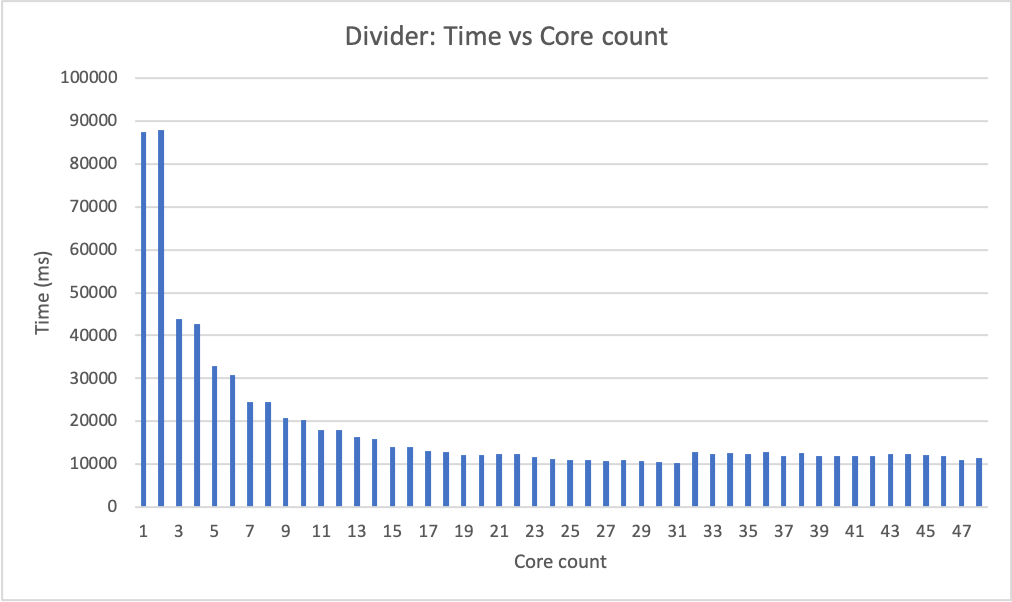}}
    \caption{Time vs Core count}
    \label{f:divcore}
\end{figure}

\paragraph{Divider.} Figure~\ref{f:divcore} shows the 16-bit divider's compute time with respect to core count. A 16-bit divider requires 32 cores to achieve the best parallel computing performance. From the chart, divider's latency decreases as the core count increases from 1 to 32. From 32 cores on, the divider's latency stays at around 12000ms. The 16-bit divider has a maximum parallel computing speedup of 8.43x, reached when it can draw on 32 cores.

\begin{figure} [ht]
    \centering {\includegraphics[width=0.8\textwidth]{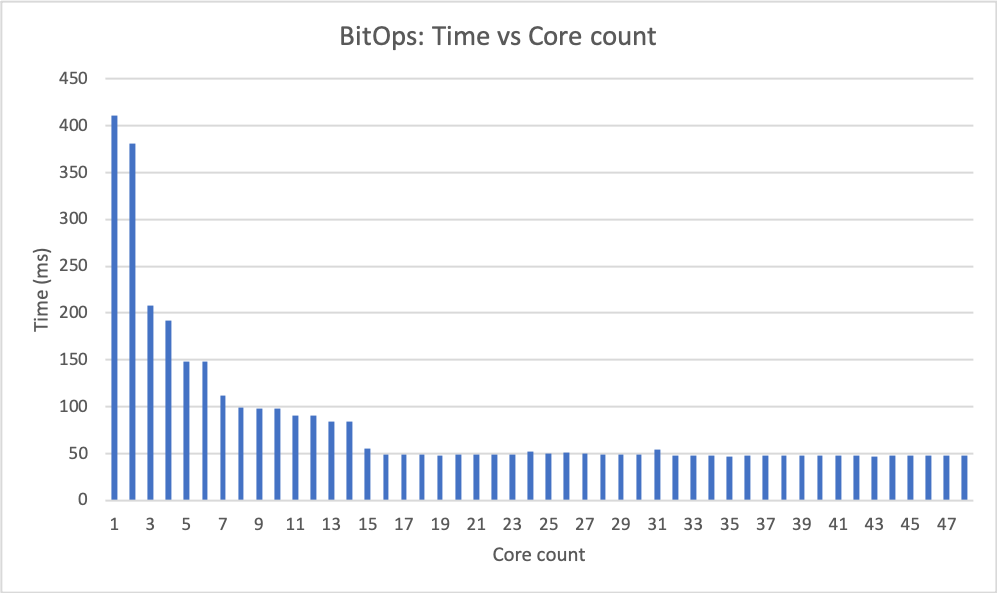}}
    \caption{Time vs Core count}
    \label{f:bitcore}
\end{figure}

\paragraph{Bitwise operations.} Figure~\ref{f:bitcore} shows 16-bit bitwise operation's compute time with respect to the core count. 16-bit bitwise ops requires 16 cores to have the best parallel computing performance. From the graph, at core count 16 the bitwise operation reached its top speed. At 16 cores and beyond, the latency stays around 50ms. 16-bit bitwise op has a maximum parallel computing speedup of 8.63x.

\begin{figure} [ht]
    \centering {\includegraphics[width=0.8\textwidth]{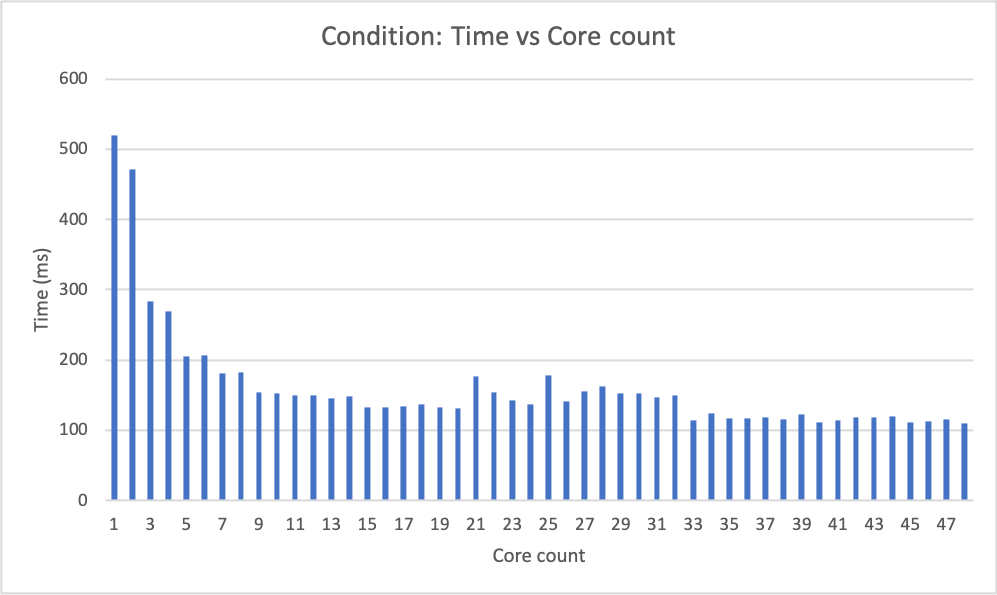}}
    \caption{Time vs Core count}
    \label{f:condicore}
\end{figure}

\paragraph{Conditional Flags.} Figure~\ref{f:condicore} shows the 16-bit conditional flag unit's compute time with respect to core count. A 16-bit conditional flag unit runs a maximum of 18 threads in parallel. On the chart from core count 15 onward, the latency of the conditional flag unit hovers at around 120-130ms. The maximum parallel computing speedup is 4.7x.

\subsection{Vulnerability}
\label{subsec:Vulnerability}
There are two source of vulnerability associated with the current CryptoEmu implementation: $(i)$ the unencrypted HE instruction memory; and $(ii)$ the branch resolution process.

Because HE instruction memory is in cleartext, the cloud knows exactly what assembly code is being executed. Assumption \ref{assumption2}  assumes that the cloud is honest, so cloud visible instructions to the cloud do not pose a security issue. However, it is possible that the cloud's execution pipeline is compromised by attackers. If an attacker obtains control over the instruction set emulator, or performs a side-channel attack, user data can be breached. 

If Assumption \ref{assumption2} is lifted, then the unencrypted HE instruction memory becomes a security loophole.Note that, we still assume the cloud will not start a denial of service attack on user. For example, if user send an encrypted number such as age, credit score or salaries, the cloud can also use binary search to actively query the value of an encrypted data using conditional unit implemented in CryptoEmu, and eventually find out the cleartext. User can counter the attack by sending HE instructions with anti-disassembly techniques, making it difficult for side-channel attack. However, this approach does not eliminate an active attack from cloud. 

The cloud is able to actively find encrypted data's value because the cloud can abuse the branch resolution process. The problem can be solved by having the user assume control over the branch resolution process \cite{irena2018cryptoblaze}. The user can ask for encrypted data from cloud, decrypt and process the data, send data back to cloud and inform the cloud which is the next instruction to fetch. With some extra overhead in data transmission, the user can make branch resolution unknown to the cloud and has full control over branch resolution.

\subsection{Comparison against the state of the art}
\label{subsec:Comparison against the state of the art}
HELib \cite{HELib2020} is one of the most widely adopted FHE software libraries. HELib implements the Brakerski-Gentry-Vaikuntanathan (BGV) scheme and supports basic homomorphic operations \cite{halevi2014algorithms}. HELib has native support for binary negation, addition/subtraction, multiplication, binary condition, shift immediate, bitwise operation and binary comparison. CryptoEmu supports all these operation and a few more: left shift/right shift with encrypted amount, unsigned binary division, and NZCV condition flag computation. This section compares the performance of the operations supported by both HELib and CryptoEmu.  

\begin{table}[ht]
\begin{center}
\begin{tabular}{lllll}
 &  & CryptoEmu & CryptoEmu& \\ 
Operation & HELib & (single core)  &  (multi-core)  & Speedup\\
\hline
Addition & 10484.1ms & 4400.67ms & 566.149ms &18.518x\\
Subtraction & 10962.2ms & 5088.57ms & 599.396ms & 18.289x\\
Multiplication (unsigned) & 69988.9ms &  86389.8ms & 10396.1ms & 6.732x\\
Multiplication (signed) & 81707.2ms & 92408.4ms & 10985.4ms &  7.438x\\
LLS (immediate)& 0.534724ms & 0.0040335ms & 0.0040335ms &  132.571x\\
Bitwise XOR & 1.52444ms  & 416.013 ms & 47.1986ms & -30.9613x\\
Bitwise OR & 771.12ms  & 416.146 ms & 47.6171ms & 16.194x\\
Bitwise AND & 756.641ms  & 411.014ms & 47.6001ms & 15.90x\\
Bitwise NOT & 1.8975ms  & 0.012508 ms & 0.012508 ms & 151.703x\\
Comparison & 4706.07ms & 519.757ms & 110.416ms & 42.62x\\

\end{tabular}
\end{center}
\caption{HELib vs CryptoEmu}
\label{table:comparison}
\end{table}

As reported in Table \ref{table:comparison}, for addition and subtraction, CryptoEmu's single core performance is about two times faster than HELib's. With sufficient cores, CryptoEmu yields maximum 18x speed up compare to HELib.
For both signed and unsigned multiplication, CryptoEmu in single core mode is slightly slower than HELib. CryptoEmu yields a maximum 7x speed up in multiple core mode compare to HELib.

HELib supports logic left shift (LLS) with immediate. This operation is not computationally intensive compared to addition/subtraction. LLS with immediate is not implemented without parallel computing optimizations on CryptoEmu, and therefore single core latency is the same as multi-core latency. Because of the excellent support provided by TFHE library, LLS with immediate on CryptoEmu is about 132x faster than on HELib.

For bitwise operation, HELib's bitwise XOR is in fact 31x faster than CryptoEmu. This is expected because the implementation for HELib's bitwise XOR is not computationally expensive. For bitwise OR and bitwise AND, CryptoEmu's single core speed is faster than HELib's. When maximum cores are enabled, CryptoEmu yields 16x speed up on bitwise OR and bitwise AND compared to HELib.

Bitwise NOT is not a computationally expensive operation on both CryptoEmu and HELib. Because bitwise NOT is implemented without parallel computing techniques, single core and multi-core performance are the same on CryptoEmu. By comparison, CryptoEmu is 151x faster than HELib on bitwise NOT operation.

For comparison/conditional operations, we benchmarked HELib's \emph{CompareTwo()} function and CryptoEmu's conditional unit. HELib's compare routine compares two encrypted values and returns the greater/lesser number and comparison result. Our CryptoEmu compares two encrypted data and returns N, Z, C, and V flags. Because the Z-flag computation is on conditional unit's critical path, it is justified to compare latencies of two functionally equivalent operations that are on both unit's critical paths. CryptoEmu's comparison is much faster than HELib's in single core mode. When multiple cores are enabled, CryptoEmu yields maximum speedup of 42x. 

\section{Conclusion and future work}
\label{sec: Conclusion and future work}

CryptoEmu successfully supports multiple homomorphic instructions on a multi-core host. With function units built upon the TFHE bootstrapped gate library, CryptoEmu is reconfigurable and reusable for general purpose computing. Through parallel computing algorithms that significantly reduces time complexity, CryptoEmu has a scalable implementation and achieves parallel computing speedup from 4.7x to 10.94x on functional units. Compare to HELib, CryptoEmu has maximum 18x speedup on addition/subtraction, 7x speedup on multiplication, 16x speed up on bitwise AND/OR, 42x speedup on comparison, 130x speedup on left shift with immediate, and 151x speedup on bitwise NOT.  

Scalability of CryptoEmu can be further improved. The current design only supports single instruction set emulator process that runs on multiple cores. Therefore maximum core count is bounded by number of cores on one CPU. With more cores available, CryptoEmu can draw on more parallelism through pipelining, multiple instruction issue and dynamic instruction scheduling. Another aspect that could provide further speed improvements is the use of AVX instructions.

\ifCLASSOPTIONcaptionsoff
  \newpage
\fi


\bibliographystyle{ieeetr}
\bibliography{ProjectCryptoEmu}

%

\pagebreak

\begin{IEEEbiography}[{\includegraphics[width=1in,height=1.25in,clip,keepaspectratio]{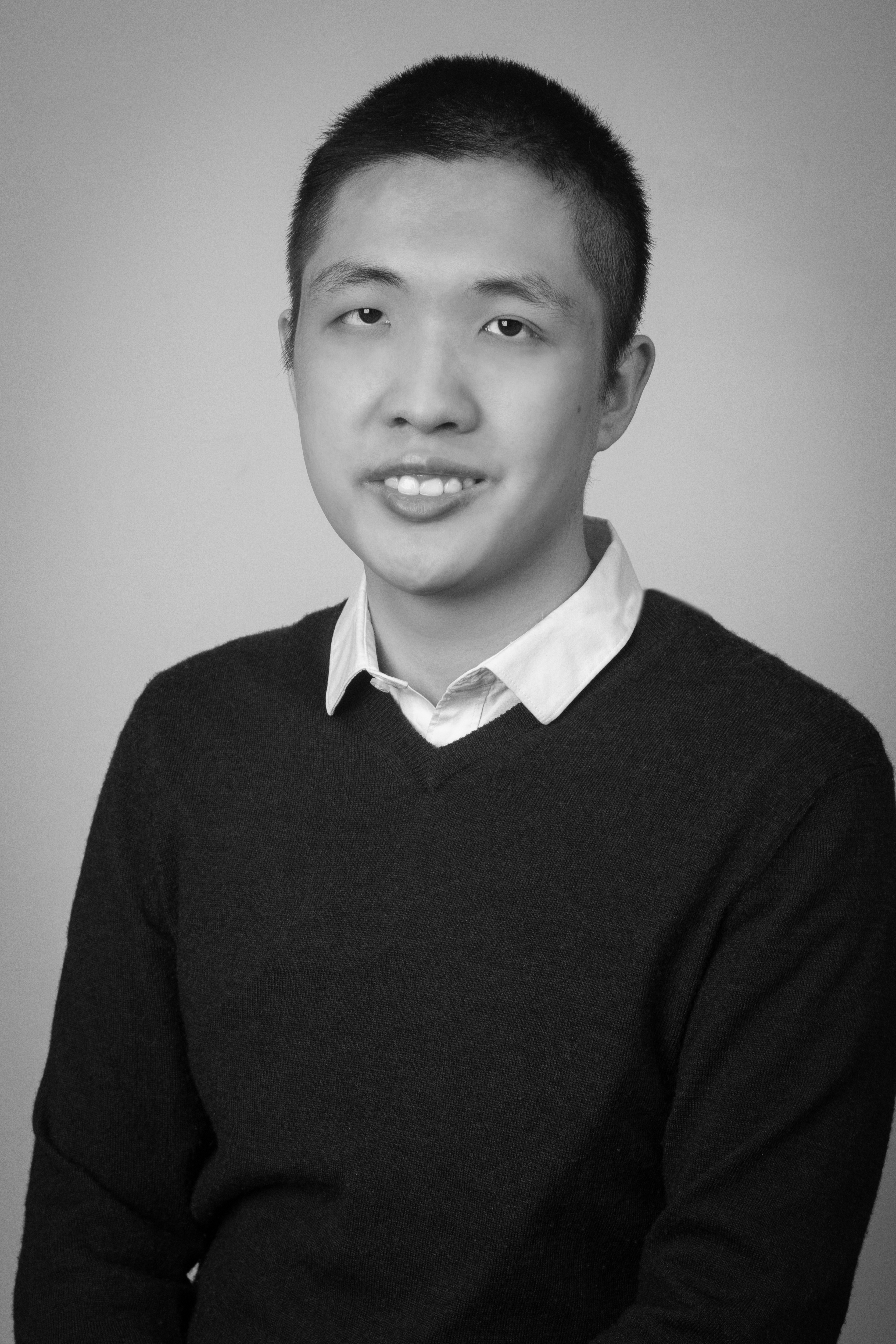}}]{Xiaoyang Gong} is a graduate student in the Department of Electrical and Computer Engineering at the University of Wisconsin-Madison. Xiaoyang received his MS in Electrical Engineering in 2020 and BS in Computer Engineering in 2019 from the University of Wisconsin-Madison. He worked as an intern at Arm Ltd.'s CPU team in 2019. His interests are in computer architecture and processor design.
\end{IEEEbiography}

\begin{IEEEbiography}[{\includegraphics[width=1in,height=1.25in,clip,keepaspectratio]{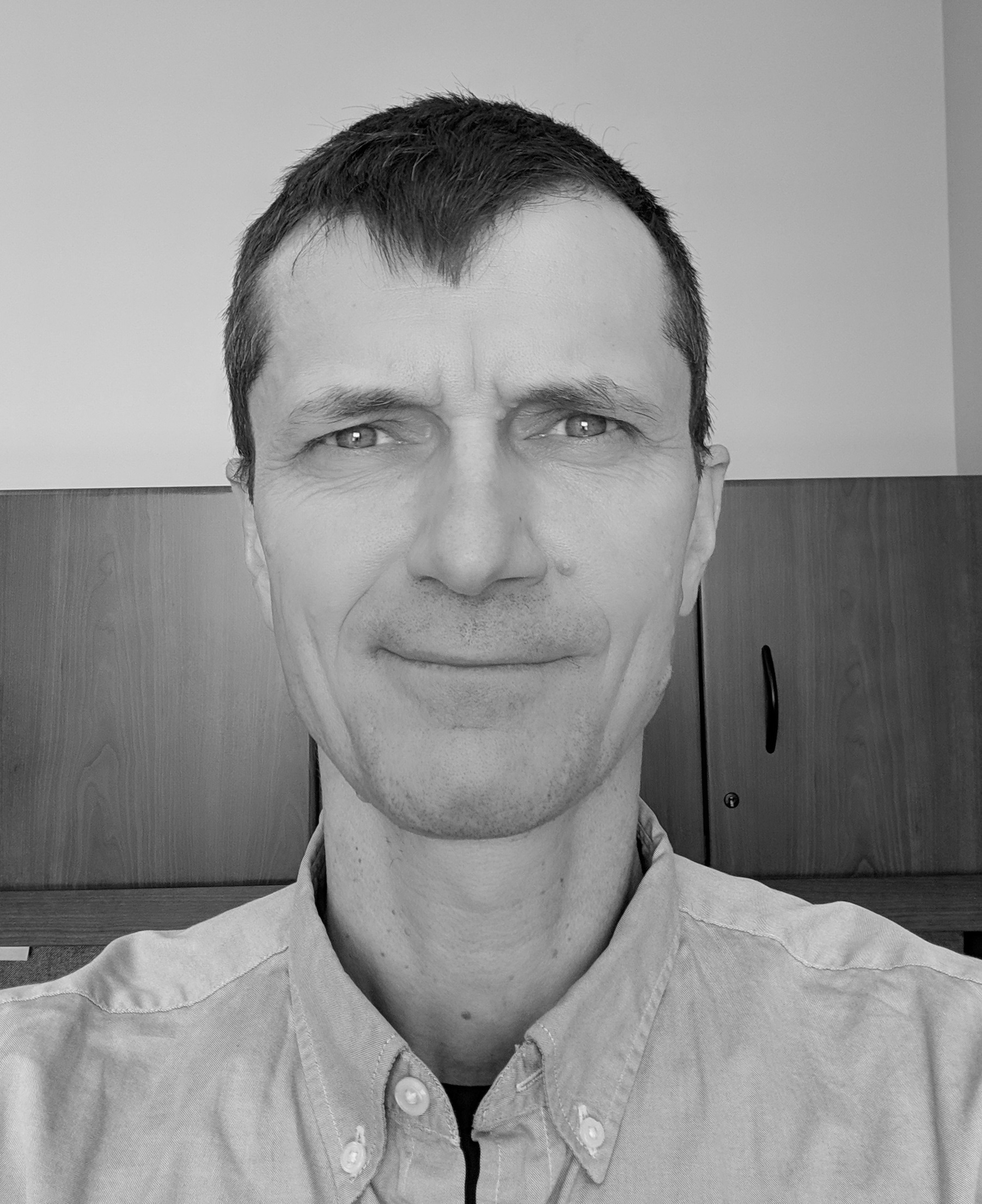}}]{Dan Negrut}
	is a Mead Witter Foundation Professor in the Department of Mechanical Engineering at the University of Wisconsin-Madison. He has courtesy appointments in the Department of Computer Sciences and the Department of Electrical and Computer Engineering. Dan received his Ph.D. in Mechanical Engineering in 1998 from the University of Iowa under the supervision of Professor Edward J. Haug. He spent six years working for Mechanical Dynamics, Inc., a software company in Ann Arbor, Michigan. In 2004 he served as an Adjunct Assistant Professor in the Department of Mathematics at the University of Michigan, Ann Arbor. He spent 2005 as a Visiting Scientist at Argonne National Laboratory in the Mathematics and Computer Science Division. He joined University of Wisconsin-Madison in 2005. His interests are in Computational Science and he leads the Simulation-Based Engineering Lab. The lab's projects focus on high performance computing, computational dynamics, artificial intelligence, terramechanics, autonomous vehicles, robotics, and fluid-solid interaction problems. Dan received the National Science Foundation Career Award in 2009. Since 2010 he is an NVIDIA CUDA Fellow. 
\end{IEEEbiography}

\end{document}